\documentclass[
 reprint,
superscriptaddress,
 amsmath,amssymb,
aps,prl
]{revtex4-2}
\usepackage{comment}
\usepackage{graphicx}
\usepackage{dcolumn}
\usepackage{bm}
\usepackage[normalem]{ulem}
\usepackage{siunitx} 
\usepackage{graphicx}
\usepackage{dcolumn}
\usepackage{bm}
\usepackage{siunitx} 
\usepackage{pgfplots}
\usepackage{tabularx}
\usepackage{soul}

\newcommand{\dd}{\mathrm{d}}
\newcommand{\ii}{\mathrm{i}}

\newcommand{\angles}[1]{\left\langle #1 \right\rangle}
\newcommand{\omg}{\omega_0}
\newcommand{\mass}{m}
\newcommand{\dx}{\varepsilon}
\newcommand{\mx}{\sigma}
\newcommand{\xmeas}{X}
\newcommand{\thermaln}{\mathrm{th}}
\newcommand{\noise}{\mathrm{N}}
\newcommand{\feedback}{\mathrm{fb}}
\newcommand{\backaction}{\mathrm{ba}}
\newcommand{\optimal}{\mathrm{opt}}

\newcommand{\gammafb}{\gamma_\feedback}
\newcommand{\gammaba}{\Gamma_\backaction}
\newcommand{\gammathermal}{\Gamma_\thermaln}
\newcommand{\optic}{\mathrm{o}}
\newcommand{\gammaoptic}{\kappa}

\newcommand{\mech}{\mathrm{m}}
\newcommand{\gammamech}{\gamma_\mech}
\newcommand{\omgmech}{\omega_0}
\newcommand{\temp}{T}
\newcommand{\kB}{k_\mathrm{B}}
\newcommand{\gammaopticcoupling}{\kappa_\mathrm{ext}}
\newcommand{\noptic}{\bar{n}_\mathrm{cav}}

\begin{document}
\preprint{APS/123-QED}
\title{Optical Interferometric Readout of a Magnetically Levitated Superconducting Microsphere}
\author{Jannek J. Hansen}
 \email{jannek.hansen@univie.ac.at}
\affiliation{University of Vienna, Faculty of Physics, Vienna Center for Quantum Science and Technology, A-1090 Vienna, Austria}
\affiliation{Institute for Quantum Optics and Quantum Information (IQOQI) Vienna, Austrian Academy of Sciences, Boltzmanngasse 3, 1090 Vienna, Austria}
 \author{Stefan Minniberger}
\affiliation{Institute for Quantum Optics and Quantum Information (IQOQI) Vienna, Austrian Academy of Sciences, Boltzmanngasse 3, 1090 Vienna, Austria}
 \author{Dominik Ilk}
\affiliation{University of Vienna, Faculty of Physics, Vienna Center for Quantum Science and Technology, A-1090 Vienna, Austria}
 \author{Peter Asenbaum}
\affiliation{Institute for Quantum Optics and Quantum Information (IQOQI) Vienna, Austrian Academy of Sciences, Boltzmanngasse 3, 1090 Vienna, Austria}
 \author{Gerard Higgins}
 \affiliation{Institute for Quantum Optics and Quantum Information (IQOQI) Vienna, Austrian Academy of Sciences, Boltzmanngasse 3, 1090 Vienna, Austria}
 \affiliation{Institute of High Energy Physics, Austrian Academy of Sciences, A-1010 Vienna, Austria}
\author{Rhys G. Povey}
 \affiliation{Institute for Quantum Optics and Quantum Information (IQOQI) Vienna, Austrian Academy of Sciences, Boltzmanngasse 3, 1090 Vienna, Austria}
 \author{Philip Schmidt}
\affiliation{Institute for Quantum Optics and Quantum Information (IQOQI) Vienna, Austrian Academy of Sciences, Boltzmanngasse 3, 1090 Vienna, Austria}
 \author{Joachim Hofer}
\affiliation{Institute for Quantum Optics and Quantum Information (IQOQI) Vienna, Austrian Academy of Sciences, Boltzmanngasse 3, 1090 Vienna, Austria}
\author{Rémi Claessen}
\affiliation{University of Vienna, Faculty of Physics, Vienna Center for Quantum Science and Technology, A-1090 Vienna, Austria}
\affiliation{Institute for Quantum Optics and Quantum Information (IQOQI) Vienna, Austrian Academy of Sciences, Boltzmanngasse 3, 1090 Vienna, Austria}
\author{Markus Aspelmeyer}
\affiliation{University of Vienna, Faculty of Physics, Vienna Center for Quantum Science and Technology, A-1090 Vienna, Austria}
\affiliation{Institute for Quantum Optics and Quantum Information (IQOQI) Vienna, Austrian Academy of Sciences, Boltzmanngasse 3, 1090 Vienna, Austria}
\author{Michael Trupke}
\email{michael.trupke@oeaw.ac.at}
\affiliation{Institute for Quantum Optics and Quantum Information (IQOQI) Vienna, Austrian Academy of Sciences, Boltzmanngasse 3, 1090 Vienna, Austria}
\date{\today}
             
\begin{abstract}
We probe the motion of a $\SI{6}{\micro\gram}$ magnetically levitated superconducting microsphere using optical interferometry at $\SI{3}{\kelvin}$,
achieving a resolution better than $\SI{1}{\nano\meter\per\sqrt{\hertz}}$, and use the measured signal to feedback-cool its motion.
The resolution exceeds the shot-noise limit of $\SI{11}{\pico\meter\per\sqrt{\hertz}}$ primarily due to technical noise arising from the roughness of the particle.
Combined with established techniques of cavity optomechanics, the high degree of isolation from environmental noise afforded by this platform provides a path to quantum physics experiments with cryogenic isolated masses at the microgram scale.
\end{abstract}
\maketitle

Mechanical quantum systems are of interest for the exploration of quantum phenomena in progressively larger and more massive systems, thereby pushing the boundaries of testing quantum mechanics \cite{Connell_2010, Palomaki_2013, Wollman_2015, Riedinger_2018, Ocheloen-Korppi_2018, Bild_2023, Galinskiy_2024, Omahen_2025}. 
Compared to clamped mechanical oscillators, levitated systems offer unique advantages, such as highly controllable trapping potentials that can, for example, facilitate sophisticated protocols in non-harmonic potentials \cite{Neumeier_2024, Roda_2024}.
In particular, significant strides have been made with optically levitated nanoparticles by demonstrating quantum control over the centre-of-mass motion \cite{Delic_2020, Magrini_2021, Tebbenjohanns_2021, Ranfagni_2022, Kamba_2023, Piotrowski_2023},
and electrical levitation has recently demonstrated mechanical quality factors exceeding $10^{10}$ \cite{Dania_2024}.
Here, we use all-superconducting magnetic levitation, which allows to trap massive particles and provides virtually dissipation-free trapping \cite{Goodkind_1999, File_1963, Romero_2012, Devlin_2019}.
Recent experiments have demonstrated the levitation of microgram masses with mechanical quality factors reaching $2.6\times 10^7$ \cite{Hofer_23}.

Precise readout of the levitator motion is a critical element on the path towards quantum experiments. 
Previously, superconducting quantum interference devices (SQUIDs) have been used to track the particle motion indirectly through its effect on the magnetic field, although the quantum-limited detection remains challenging \cite{vanWaarde2016, Gutierrez_2023, Hofer_23, Schmidt_2024}. 
Optical interferometers are the most precise measurement tools for relative position changes \cite{Abbott_2016}, promising the resolution required to enable quantum control of magnetically levitated particles.
However, they are incompatible with superconductors because the single-photon energy is above typical Cooper-pair binding energies.
Lately, progress has been made to combine optical photons with superconducting circuits \cite{Delaney_2022, Meesala_2024, Arnold_2025}. In this letter, we successfully demonstrate the direct optical readout of a levitated superconductor.

\begin{figure}[!tb]
    \centering\includegraphics[width=1\linewidth]{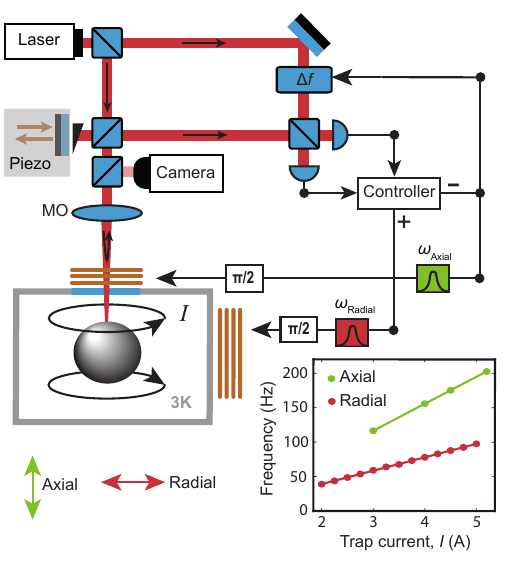}
    \caption{
    Schematic of the setup.
    We levitate a superconducting microsphere in the magnetic field minimum between anti-Helmholtz coils.
    The axial position of the particle is measured by interferometry. We lock the phase with an acousto-optic modulator ($\Delta f$) to increase the linear dynamic range of the interferometric readout. One arm of the interferometer is focused onto the particle by a microscope objective (MO).
    The sum of the detectors output contains intensity information which is used to feedback the radial motion, while the difference contains phase information corresponding mainly to the axial motion. The feedback is applied by bandpass filtering around the corresponding frequency and adding a phase delay to the measured signal. The remaining signal is applied to additional wires outside the cryostat (brown) through a current amplifier.
    The interferometer signal can be calibrated using a mirror mounted on a piezoelectric actuator (grey).
    We also record the particle position using a camera.
    \textit{Inset:} The motional frequencies are linearly dependent on the trap current.
    }\label{fig:setup}
\end{figure}

Our experimental setup is shown schematically in Fig.~\ref{fig:setup}.
We levitate a $\SI{6}{\micro\gram}$ lead-tin (PbSn) superconducting sphere of radius $\SI{50}{\micro\meter}$ in an anti-Helmholtz magnetic trap made from coils of niobium-titanium wire within a $\SI{3}{K}$ closed-cycle cryostat.
The particle oscillates about the trap centre with frequencies
$f_i=\sqrt{3/(8\,\pi^2\,\mu_0\,\rho)}\,b_i$ \cite{Romero_2012}, where
$\mu_0$ is the permeability of free space, $\rho$ is the particle density, and $b_i$ is the gradient along the $i$ direction that is linearly dependent on the coil current.
In the azimuthally symmetric anti-Helmholtz trap, the axial field gradient $b_z$ and the axial frequency $f_z$ are double the radial-plane field gradients $b_x, b_y$ and frequencies $f_x, f_y$ \cite{Hofer_2019}.
Optical interferometry along the axial z-direction is performed with the particle as one of the two mirrors in a Mach-Zehnder-like interferometer, using a $\lambda=\SI{637}{\nano\meter}$ laser and balanced homodyne detection.
Because radial motion of the particle affects the amount of light collected at the detectors, we pre-cool the particle with camera-based and laser-intensity-based methods that take advantage of the axial optical access.
Further details on the experimental setup are available in the supplementary material \cite{Suppl}.

An experimental run begins with the superconducting microsphere resting on a 
silicon carbide (SiC) substrate at the cryostat base temperature of $\SI{3}{\kelvin}$.
We start by applying a large current through the magnetic trap coils to overcome any adhesion forces and lift the particle off the substrate. To reduce residual axial motion, we use friction. Specifically, we briefly lower the current after lift off, such that the microsphere returns to the SiC surface, where it is frictionally damped through surface contact. The trap current is then slowly ramped up again until the particle is placed into the centre of the trap. 
Initially, the particle moves with a large radial amplitude of around $\SI{50}{\micro\meter}$,
and is feedback cooled using external magnetic fields generated with dedicated wires outside the cryostat.

The first cooling step uses a camera-based readout. We illuminate the particle with an LED flash and image the reflected light through the microscope objective.
Distances are calibrated using the known microsphere diameter.
Using two camera snapshots separated by 20\% of the radial period, we discern the particle motion and apply a counteracting magnetic pulse through the external feedback wires.
Repeating this process, we reduce the radial amplitude to around $\SI{200}{\nano\meter}$.

\begin{figure}
    \centering\includegraphics[width=1\linewidth]{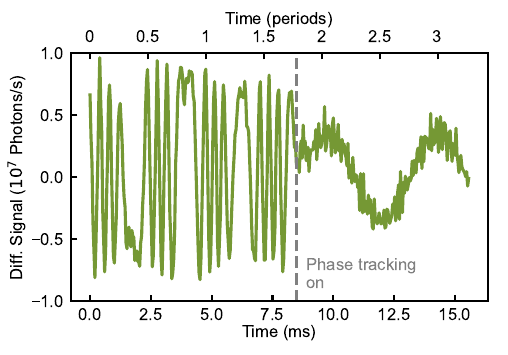}
    \caption{
    Linearisation of the interferometric signal with balanced homodyne detection. Throughout this time trace the particle motional amplitude is approximately $\lambda$, resulting in multiple oscillations in the time trace.
    Linearisation by phase tracking allows the extraction of a sinusoidal signal proportional to the particle displacement.
    }
    \label{fig:linearisation}
\end{figure}

Next, laser-intensity based measurements are used to continuously stabilize the particle position.
A piezoelectric stage holding the microscope objective is used to position the laser beam onto the particle.
Because the microsphere is curved, the angle of the reflected light, and hence the collection efficiency, depends on the radial displacement.
We position the laser beam slightly off-centre, such that radial motion of the particle induces a power modulation of the detected light.
The summed signal of the photodetectors, corresponding to intensity, is then bandpass-filtered and phase-shifted to supply a current to the radial feedback wires. The current creates an oscillating magnetic field that opposes the radial motion of the particle.
Finally, the difference signal of the photodetectors is used for interferometric readout.
A conventional interferometer responds linearly only to phase shifts far below $\pi/2$ \cite{Luiz_2019}, corresponding to a displacement of $\lambda/{8}$ in our setup.
To increase the dynamic range of the interferometer to displacements of several $\lambda$, we employ a high-gain closed-loop feedback \cite{Fisher_1979}. This method uses an acousto-optic modulator (AOM) to track the frequency difference between the signal and reference arms of the interferometer.
Fig.~\ref{fig:linearisation} compares the interferometer signals with and without phase tracking.
When the lock is active, the signal is linearly proportional to the particle axial displacement, with the constant of proportionality selected to achieve the desired dynamic range.
Cooling the axial motion is performed in the same way as above.

Two independent calibration methods are used to convert the differential signal from the photodiodes to absolute displacement.
First, we reflect light from a mirror mounted on a piezoelectric stage (Fig.~\ref{fig:setup}) and measure the interferometer signal when we drive oscillatory motion of the mirror with a known amplitude.
Second, we apply an oscillating magnetic field on the particle using the axial external feedback wire, and measure the particle response when different trapping frequencies (Fig.~\ref{fig:ForceCallibration}).
More details on the calibration methods are given in the supplementary material \cite{Suppl}.

We measure a calibrated axial-displacement-amplitude one-sided spectral density noise floor of $\sqrt{2\,S_{zz}} = \SI{955\pm148}{\pico\meter\per\sqrt{\hertz}}$ near the mechanical frequency.
Ideally, our precision would be limited by the photon shot noise ${S_{zz}^\mathrm{shot}}=\lambda^2 / (64\,\pi^2\,n_\mathrm{det})$, where $\lambda$ is the laser wavelength, and $n_\mathrm{det}$ is the average collection rate of photons scattered off the particle \cite{Clerk_2010}.
With $\lambda=\SI{637}{\nano\meter}$ and collection rate $n_\mathrm{det}=10^7\,\SI{}{\text{photons}\per\second}$, this theoretical limit is $\sqrt{2\,S_{xx}^\mathrm{shot}} = \SI{11}{\pico\meter\per\sqrt{\hertz}}$.
The greater noise floor observed in our experiment is dominated by the microsphere roughness, together with rotations and drifts within the trap (see supplementary material \cite{Suppl}).

Despite direct optical readout, the particle remains superconducting up to several hundred seconds.
However, to avoid any quenching, we limit experimental run-times to $\SI{100}{\second}$ and laser-photon fluxes to $\SI{1e7}{photons/\second}$.
Between runs, we rest the particle on the $\SI{3}{\kelvin}$ substrate, allowing it to re-thermalise for $\SI{100}{\second}$.
Further discussion of the trapping lifetime is provided in the supplementary material \cite{Suppl}.

\begin{figure}
    \centering
    \includegraphics[width=1\linewidth]{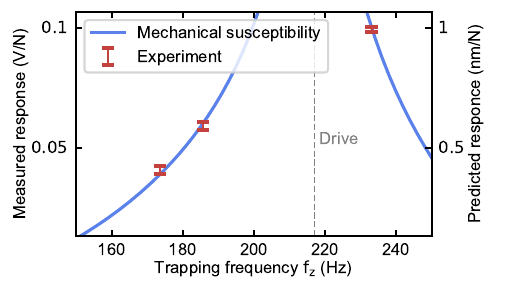}
    \caption{Calibration of the interferometer signal. An oscillating magnetic force is applied at \SI{217}{Hz} with an external coil. The response of the particle is measured for three trapping frequencies, $\{174,186,233\}\SI{}{\hertz}$. A careful characterisation of the applied force allows calibration of displacement units (right axis) from the measured response (left axis). 
    }
    \label{fig:ForceCallibration}
\end{figure}

\begin{figure}
    \centering
    \includegraphics[width=1\linewidth]{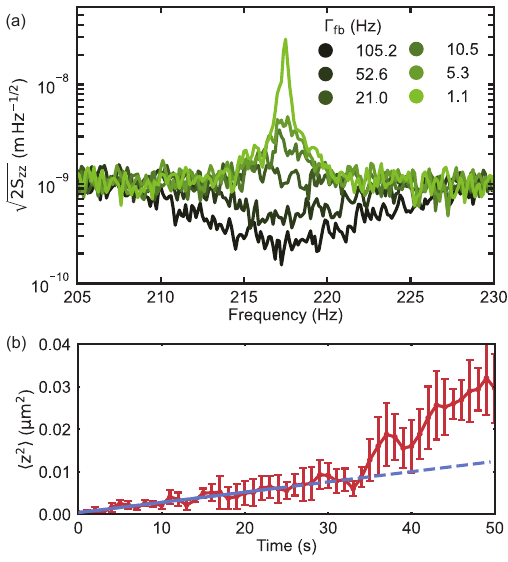}
    \caption{
    (a) Measured one-sided power spectral density, in units of displacement noise, taken with different feedback rates. Each trace is produced from the Fourier transform of time series data. Increasing feedback strength (darker curves) reduces the displacement amplitude, pushing the measured response below the noise floor when the optimal feedback gain is exceeded.
    (b) Ring-up of the motional energy of the mechanical oscillator from its cooled state is used to fit the thermal decoherence rate ($\Gamma_\mathrm{th}=\SI{6.4e12}{Hz}$ at a trap frequency of \SI{160}{Hz}) at low amplitudes. At higher amplitudes (after $\sim$\SI{35}{s}), faster heating is observed, likely caused by coupling between modes due to trap anharmonicity.
    }
    \label{fig:squashing}
\end{figure}

Cooling dynamics for interferometric feedback on the axial motion are shown in Fig.~\ref{fig:squashing}. With moderate gain, the motion of the particle is cooled to an amplitude below $1\,\lambda$ to be in the linear regime of the detection. By increasing the feedback gain, the amplitude is further reduced. Optimal feedback cooling, corresponding to a flat spectral density, is found around $\Gamma_\mathrm{fb}^\mathrm{opt}=\SI{21}{\hertz}$, and gives a minimum cooled displacement oscillation of $\sqrt{\langle z^2 \rangle_\mathrm{min}} \approx \SI{8}{\nano\meter}$ corresponding to mechanical phonon occupation of around $\Bar{n}\approx 5\times10^{12}$ and a temperature equilibrium of \SI{52e3}{K}. 
For high gains, the particle dynamics are dominated by the fed-back measurement noise, and classical noise squashing \cite{Poggio_2007,Magrini_2021} below the measurement noise floor can be observed.
The initial effective environmental temperature of the levitator motion is significantly above the thermal temperature due to vibrational noise.

After the motional amplitude of the levitator is cooled, its energy is below that of the environmental bath dominated by mechanical vibrations.
When cooling is stopped, the levitator motion is heated by external excitations mediated by the mechanical damping rate $\gamma=\omega_0/Q$ (Fig.~\ref{fig:squashing}b). The linear increase of the particle energy 
can be expressed by $n_0 + n_\mathrm{th}\,\gamma\,t = n_0 + \Gamma_\mathrm{th}\,t$ where $n_0$ is the initial cooled phonon occupation of the oscillator, $n_\mathrm{th}$ the thermal equilibrium phonon occupation, and $\Gamma_\mathrm{th}=n_\mathrm{th}\,\gamma$ is the thermal decoherence rate \cite{Magrini_2021, Dania_2024}. We observe a change in the rate at a certain amplitude; this behaviour has been observed in other experiments and can be related to an amplitude-dependent coupling to other modes \cite{Gutierrez_2023}.

With the measured thermal decoherence rate and the induced backaction rate $\Gamma_\mathrm{ba}$ by $10^7$ photons per second the quantum cooperativity $\mathcal{C}_\mathrm{q} = \Gamma_\mathrm{ba}/\Gamma_\mathrm{th}\approx1.4\times10^{-21}$ of the system is far from the required value to enable feedback-based quantum control of the oscillator.

We now outline a path to cooling a magnetically levitated microgram object to its mechanical quantum ground state using measurement-based feedback. This method requires high resolution of the particle position, which implies a high collection and detection efficiency, $\eta$ \cite{Magrini_2021, Tebbenjohanns_2021}.
Even with $\eta=1$, and a quality factor of $2.6\times10^7$ at $\SI{3}{\kelvin}$, our system would require at least $7.5\times10^{17}\,\SI{}{\text{photons}\per\second}$ or $\SI{235}{\milli\watt}$ to sufficiently suppress the shot noise for ground state cooling. 
This is practically unfeasible due to associated heating, but can be avoided by implementing an optical cavity \cite{Aspelmeyer_2014, Abbott_2016, Ernzer_2023, Melo_2025} where the phase information acquired by each photon is enhanced so that fewer photons are required to probe the mechanics with the same precision.

\begin{figure}
    \centering
    \includegraphics[width=1\linewidth]
    {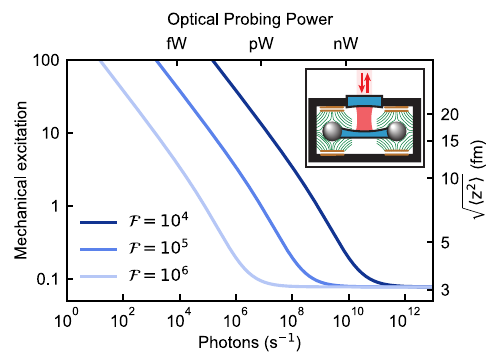}
    \caption{Simulated number of incoming optical photons needed to cool a mirror levitator assuming optimal feedback. System parameters given in the main text with varying optical cavity finesse, $\mathcal{F}$.
    The number of incoming photons required to reach the desired mechanical excitation depends inversely on the square of the finesse.
    \textit{Inset:} Superconductors are used to levitate a mirror which forms part of an optical cavity.
    }
    \label{fig:high_finesse_cavity}
\end{figure}

An implementation of this concept is shown in the inset of Fig.~\ref{fig:high_finesse_cavity}, where we consider the levitator as one of the mirrors in an optical cavity.
Describing the optical cavity with resonant angular frequency $\omega_\mathrm{cav}$, power loss rate $\kappa$, and average occupation $n_\mathrm{cav}$,
the optomechanical system has
$\mathcal{C}_\mathrm{q}=\mathcal{C}_\mathrm{om}/n_\mathrm{th}$, where $\mathcal{C}_\mathrm{om}=4\,g^2\,n_\mathrm{cav}/(\kappa\,\gamma)$ is the optomechanical cooperativity and $g=z_\mathrm{zpf}\,\delta\omega_\mathrm{cav}/\delta z$ is the (bare) optomechanical coupling. The minimum requirement to cool to a mechanical occupation below 1 phonon is given by: 

\begin{align*}
n<1\quad\Rightarrow\quad
\frac{4\,g^2\,n_\mathrm{cav}}{\kappa\,\Gamma_\mathrm{th}} > \frac{1}{9\,\eta-1}>0 \;,
\end{align*}
where $\eta = n_\mathrm{det}/n_\mathrm{in}$ is the efficiency. Cooling to the ground state thus requires a minimum number of photons, for a one-sided over-coupled optical cavity ($n_\mathrm{cav}=4\,n_\mathrm{in}/\kappa$) at $\lambda=\SI{1.55}{\micro\meter}$,
finesse $\mathcal{F} = 10^5$, and $\eta = 0.75$, with $m = \SI{6}{\micro\gram}$, $f = \SI{200}{\hertz}$ and $Q = 2.6\times10^7$ quality factor, temperature $T = \SI{15}{\milli\kelvin}$, ground state cooling needs an input of only $n_\mathrm{in}=\SI{7E6}{\text{photons}\per\second}$ or $\SI{0.9}{\pico\watt}$.
In Fig.~\ref{fig:high_finesse_cavity} we plot this requirement, showing the mechanical excitation as a function of input photons, assuming optimal feedback cooling, for varying optical cavity finesse.
The measurement noise beyond shot noise $S_{zz}^{\mathrm{imp}+}$ needs to be below $\sqrt{2\,S_{zz}^{\mathrm{imp}+}}<\SI{0.8}{\femto\meter\per\sqrt{\hertz}}$ for our parameters, similar system parameters have already been achieved \cite{Hunger_2010, Muller_2010, Fait_2021, Abdelatief_2025}. 
An overview of optomechanical feedback cooling is provided in the supplementary material \cite{Suppl}.
Other improvements, such as higher mechanical frequencies by using ring- or disk-shaped superconductor levitators \cite{Gutierrez_2020, Navau_2021, Hofer_2024, Bort_2024} are also possible.

In summary, we precisely read out the motion of a levitated superconducting microsphere using optical interferometry, with a sensitivity of around $\SI{1}{\nano\meter/\sqrt{\hertz}}$. Modifications to the system, in particular the integration of high-finesse cavity readout, will allow cooling of the particle to the quantum mechanical ground state of its centre-of-mass motion. The high resolution and isolation from the surroundings make this platform a promising path towards quantum experiments with masses in the microgram regime, eventually enabling the exploration of gravitational quantum phenomena \cite{DeWittRickles_2011, Pino_2018, Krisnanda_2020, Bose_2025}. 
\\
The data of this study are available at the Zenodo repository \cite{data_hansen_2025}.

\textit{Acknowledgements:} We gratefully acknowledge valuable discussions with U. Deli\'{c}, P. Treussart, P. Koller, S. Putz, and L. Magrini. This work was supported by the European Union’s Horizon 2020 research and innovation program under Grant No. 101080143 (SuperMeQ) and the European Research Council under Grant No. 951234 (ERC Synergy QXtreme). This research was funded in whole or in part by the Austrian Science Fund (FWF) [10.55776/ESP525]. For open access purposes, the author has applied a CC BY public copyright license to any author-accepted manuscript version arising from this submission.

\newpage

\section{Supplemental Material}
\subsection{Experimental setup}

\begin{figure}[b!]
    \centering
    \includegraphics[width=0.8\linewidth]{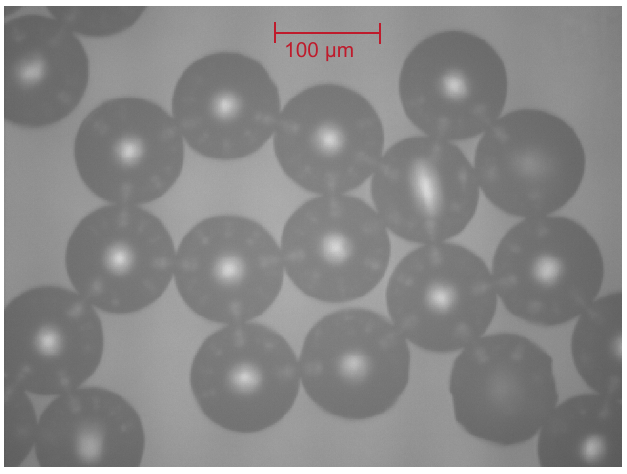}
    \caption{Microscope image of Sn10Pb90 solder balls.}
    \label{fig:Sphers}
\end{figure}

\begin{figure*}[t!]
    \centering
    \includegraphics[width=0.8\linewidth]{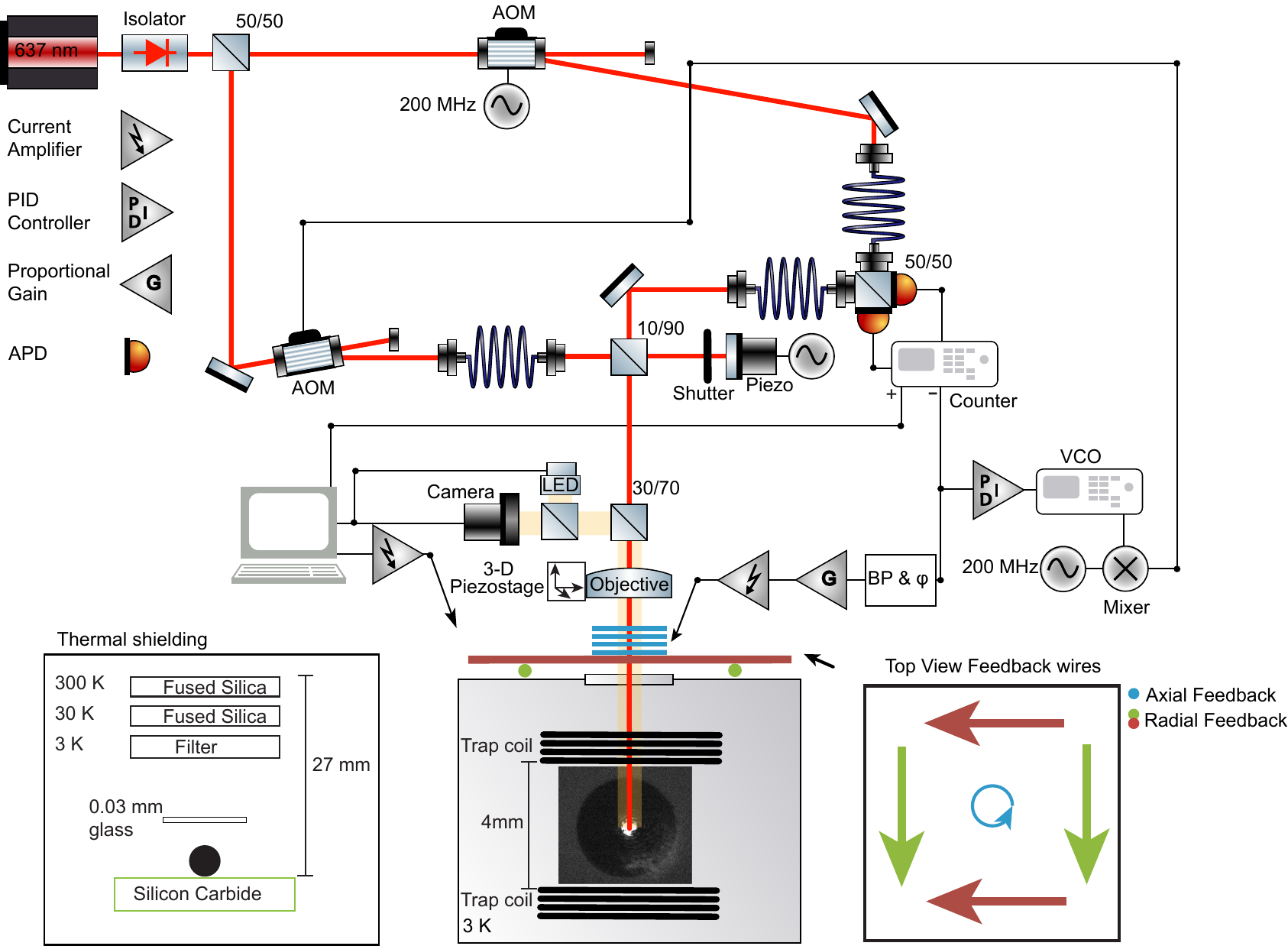}
    \caption{Schematic of the experimental setup. A diode laser supplies coherent light with a wavelength of \SI{637}{nm} to a 50:50 beam splitter, which separates the light into two paths, creating a Mach-Zehnder-like interferometer. For the linearization lock described in the text, the laser is modulated in both the experiment path and the local oscillator path by two independent acousto-optic modulators (AOMs), which shift the frequency difference between the two arms. The laser beams are recombined and directed onto the top of the levitated particle using a wavelength-correcting microscope objective (MO) mounted on a 3D piezo stage with a travel range of \SI{20}{\micro\meter} in all axes. 
    Levitation of the superconducting sphere is carried out with a superconducting anti-Helmholtz coil-pair magnetic trap with up to \SI{6}{A} of current inside a closed-cycle cryostat at \SI{3}{K}. A camera is placed above the objective to image the radial position of the particle. The objective collimates light reflected by the particle. Here, we lose 30\% at the camera beam splitter and 10 \% at the beam splitter redirecting the light into the Mach-Zehnder interferometer. The remaining 63\% are recombining with the local oscillator at the 50:50 beam splitter. The outputs of the beam splitter are connected to two single-photon counting modules (APD), creating a balanced homodyne detection scheme. The counts are converted to a proportional voltage signal by an FPGA-based counter. The difference in the photon counts is used to feed back on the motion of the particle in the axial direction, while the sum of the photon counts carries information about the radial motion. The sum is passed on to a computer running a \textit{LabView} program, calculating a feedback signal at the radial frequency and passing it onto a current amplifier. The differential signal is passed on to a PID controller, which creates an error signal controlling a voltage-controlled oscillator (VCO) whose output is mixed onto the AOM drive in the signal arm. When the signal in the differential output is linearized, it is passed onto an FPGA-based bandpass filter and passed with a time delay and gain onto a current amplifier driving the axial feedback coil (blue). The thermal shielding consists of a fused silica window at \SI{30}{K}, an optical bandpass filter thermalized to the \SI{3}{K} stage and a narrow opening to the particle holder. Additionally, a \SI{30}{\micro \meter} glass plate is placed on the holder, preventing the particle from escaping the housing. On top of the vacuum chamber, copper wires are mounted in two counter-propagating coils to exert a horizontal force on the particle (green: x-axis, red: y-axis). For the axial control, a radial coil is placed directly above the particle (blue).}
    \label{fig:Setup}
\end{figure*}

The superconducting sphere is levitated inside a \textit{Montana Cryostation} with a high numerical aperture kit, which reaches a temperature of \SI{3}{K} at the experimental platform. The particle is kept in an enclosed housing, where the bottom is made of silicon carbide (SiC) for its high thermal conductivity \cite{Cheng_2022}, and the top is covered by a \SI{30}{\micro m} thick glass lid to prevent escape. 
To minimize room temperature radiation from heating the particle, the setup has a fused silica window at the \SI{30}{K} stage and an optical bandpass filter at the \SI{3}{K} stage. 
The trap coils are made from copper-matrix niobium titanium (NbTi) wire, wound around an aluminum holder in an anti-Helmholtz configuration. A voltage-controlled current supply (\textit{Delta Elektronika}), connected via a high current feedthrough, supplies the trap with up to \SI{6}{A}. Conceptually, the trap is comparable to the one presented in \cite{Hofer_2023}, but without vibration isolation.

The particle is a commercial Sn10Pb90 (10\% tin, 90\% lead, by weight) solder ball from \textit{Easyspheres}, with a specified diameter of 100 $\pm$ \SI{6}{\micro m}.
The displacement resolution reached in this experiment exceeds the shot-noise limit by two orders of magnitude. We ascribe this discrepancy to surface roughness and shape imperfections of the levitated PbSn particle, which lead to intensity noise and phase fluctuations as the particle rotates in the trap. A microscope image of a sample of solder balls is shown in Fig.~\ref{fig:Sphers}. Shape imperfections are clearly discernible. A white light interferometer scan of the surface of a solder ball was performed (not shown), giving a surface roughness of $\sigma_r  =\SI{50}{\nano \meter}$ over a region of \SI{10}{\micro\meter}$\times$\SI{10}{\micro\meter}. 

At rest on the SiC surface, the sphere thermalizes with the cryostat and is assumed to reach base temperature before liftoff. Levitation was achieved with starting temperatures ranging from \SI{3}{K} to \SI{6.4}{K}.
Once levitated, the particle can dissipate internal energy only very slowly through blackbody radiation. Heating due to illumination and stray light therefore eventually quenches the superconductor, leading to finite levitation times. An investigation of the relationship between laser power and levitation lifetime is described in Sec.~\ref{sec:lifetime}. 

Fig. \ref{fig:Setup} provides an overview of the experimental setup. Global experimental sequence control is provided by a \textit{Labview} program, which also controls the feedback loops. To image the position of the particle, a camera (\textit{The Imaging Source Europe GmbH, DMK 37BUX287 CMOS} with a resolution of 560 × 720 pixels) and an LED flashlight with an analog trigger are installed above the chamber. Laser light is provided by a \textit{Toptica DLC Pro} with a wavelength of \SI{637}{nm} (red), which is then split into two paths by a 50/50 beam splitter. Both paths are equipped with an acousto-optic modulator (AOM) to change the differential frequency between the two paths. One path represents the reference arm of the interferometer and is directly coupled to the fiber beamsplitter, whilst the other path is mode cleaned with a single-mode fiber and directed onto the particle. The alignment of the laser onto the particle is performed with a microscope objective mounted on a three-dimensional piezo stage. An intensity scan of the position, as shown in Fig.~\ref{fig:intensity_cooling}, is used to choose the location of the laser probe for an experimental run.

The reflection of light from the particle is mixed with the local oscillator arm in a 50/50 fiber beam splitter and detected by two photon-counting modules (\textit{Excelitas SpCM-AQRH-10-FC}) in a balanced homodyne configuration.
Logic voltage pulses corresponding to detected photons are converted within a time-bin to a number-proportional voltage by a field-programmable gate array (FPGA) (\textit{Liquid instruments Moku:Go}) with a rate of \SI{200}{kHz}.
Another FPGA (\textit{Moku:Go}) combines these two detector outputs into a sum and difference signal. 
For interferometric-based axial-motion cooling, the difference signal is sent to an FPGA-based bandpass filter (\textit{red Pitaya} using \textit{PyRPL} \cite{PyRPL}) to convert it to the direct axial feedback signal applied to a custom-built voltage-controlled current amplifier, and to an FPGA-based voltage-controlled oscillator (\textit{Moku:Go}) to shift the frequency of the AOM locking the interferometer.
The sum of the signal corresponds to the intensity and is used to cool the radial motion. The signal is processed by the \textit{Labview} program which controls a custom-built current supply to the radial feedback coils.

\subsection{Experimental sequence}
\begin{figure*}[t!]
    \centering
    \includegraphics[width=0.8\linewidth]{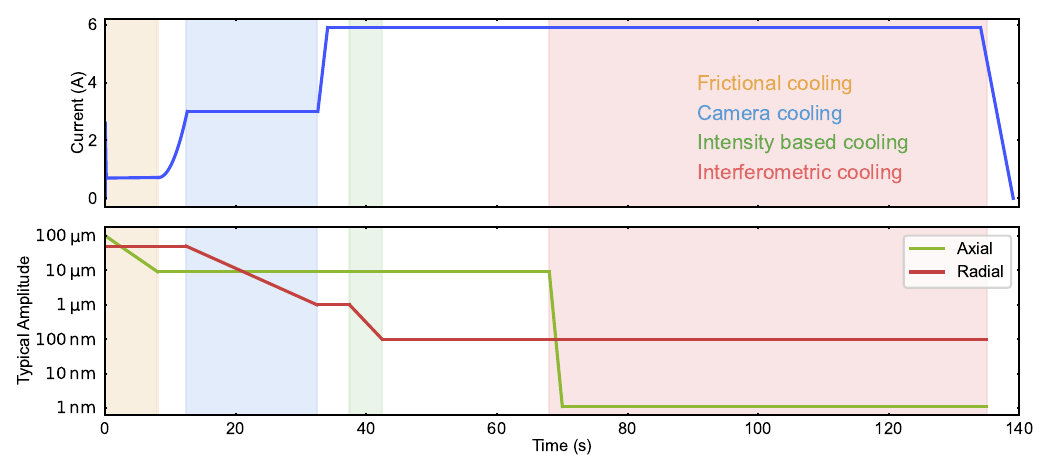}
    \caption{A typical experimental sequence. The particle is first cooled on the bottom of the trap using friction, by choosing the current to be large enough to roll the particle around on the ground. After the lift-off of the particle, the radial motion is damped using pulsed camera detection-based feedback. Then, focusing the laser onto the sphere with a known offset in the radial plane, the intensity of the reflected laser is used to suppress the radial motion of the sphere further. Finally, it is possible to recollect enough reflected light to lock the interferometer and feed back on the axial motion to the noise floor of the interferometric detector.}
    \label{fig:Sequence}
\end{figure*}

As illustrated in Fig.~\ref{fig:Sequence}, cooling of the particle is performed in 4 steps.

\subsection{Frictional cooling}
In order to get the particle into the trap with minimal excess axial motion, we start by launching the particle off the ground with a rapid ramp to \SI{2.6}{A} in \SI{0.07}{s} of the trap current, such that any adhesion forces are overcome. Then, the current is ramped down to \SI{0.7}{A} such that the sphere returns to the SiC surface but can roll across it, eventually approaching the local minimum of the magnetic field.
The trap current is then increased following a quadratic ramp to \SI{3}{A} in \SI{4.4}{s}, so as to adiabatically increase the axial trapping frequency to a value of \SI{120}{Hz}.

\subsection{Camera-based cooling}
Next, it is necessary to reduce the radial motion of the particle such that its apex remains within the laser spot of \SI{2}{\micro m} diameter.
Using the known frequency of the particle's trajectory in the trap, we apply camera-based feedback to reduce the radial amplitude. First, the center of the radial motion is found by recording a video with 30 frames.
Then, damping sequences with timed, directional magnetic pulses are applied.
This sequence begins with two images taken \SI{3}{ms} apart, followed by a waiting time of two oscillation periods to allow the computer to calculate the required feedback direction, and then the appropriate pulse is applied through the radial feedback coils.
After 25 iterations, this pulsed cooling method reduces the radial amplitude of the particle to below \SI{1}{\micro m} RMS, which is sufficient to keep the particle within the laser beam.

\subsection{Intensity feedback cooling}
\begin{figure}
    \centering
    \includegraphics[width=1\linewidth]{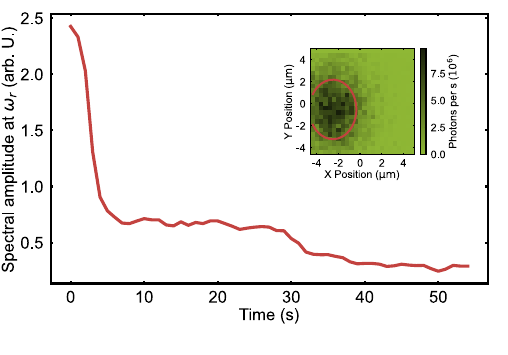}
    \caption{The graph shows the amplitude of the radial frequency component over time, while intensity-based radial feedback cooling is applied. Continuous feedback is applied to the radial frequency in the direction chosen using the intensity profile. This feedback scheme can cool the radial motion, even when the motion becomes larger than the linear regime of the intensity signal. After \SI{30}{s}, the axial motion begins cooling independently, decreasing the intensity noise such that the radial feedback limit improves. The vertical axis is calibrated to micrometers. However, this calibration only holds for small displacements along the gradient of the intensity profile around the rest position. Inset: By scanning the microscope objective in the radial plane above the levitated particle, an intensity profile can be measured depending on the particle's position. This profile is used to choose a position at an intensity slope along the axis of the radial feedback coil.}
    \label{fig:intensity_cooling}
\end{figure}

With the particle radial displacements small enough, we now position the laser beam and use the reflected intensity (summed signal) to damp the radial motion further.
To map out the sphere's location relative to the laser, the piezoelectrical controllable stage holding the focusing lens is scanned in the microscope plane, measuring the intensity at each point for multiple periods of the radial motion to obtain an average intensity.
As shown in Fig.~\ref{fig:intensity_cooling}, the intensity profile has a Gaussian shape with a full-width at half-maximum of roughly \SI{4}{\micro m}.
The lens is positioned slightly off-center on the slope of the Gaussian, such that the instantaneous variation in reflected intensity is monotonic with respect to the particle's radial displacement.
This sum signal, filtered around the radial frequency and with appropriate phase delay, is then applied to the radial feedback coils, reducing the transverse amplitude to approximately \SI{100}{\nano\meter}.

\begin{figure}
    \centering
    \includegraphics[width=1\linewidth]{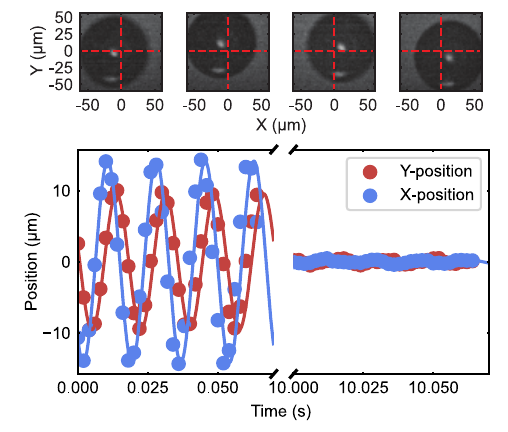}
    \caption{Upper panels: Using the camera pictures, the program determines the position of the particle in both radial coordinates. To feedback on the motion of the particle, the program finds the quadrant in which the particle is located and triggers a corresponding magnetic kick towards the centre of rotation. The four pictures show four different images of the particle, where it is respectively in one of each of the four quadrants.
    Lower panel: The radial motion of the sphere is recorded with a camera mounted above the microscope objective. Camera-based feedback reduces the motion of the sphere from tens of micrometers to less than \SI{1}{\micro\meter} rms over a period of \SI{10}{s}. The blue and red data points correspond to the particle's displacement from the trap's center in two orthogonal radial axes before and after performing the camera cooling protocol.}
    \label{fig:Camera_cooling}
\end{figure}

\subsection{Interferometric phase tracking and cooling}
We now cool the axial motion with the interferometric (difference) signal.

The axial amplitude of the particle exceeds $\lambda/8$, which corresponds to a phase shift of $\pi/2$ in the interferometer; the signal is wrapped as shown in the main text. To increase the dynamic range of the interferometer and also account for drifts and the surface roughness of the particle, we apply a strong lock on the interferometric signal. 
The lock consists of a proportional controller using the voltage signal of the difference between the two detectors. The lock controls a VCO, the output of which is mixed into the drive of the AOM in the signal arm. This frequency shift compensates the Doppler shift induced in the reflected light by the velocity of the particle. Using this method, we increase the linear regime of the signal to several \si{\micro m}. As it also suppresses the actual signal from the oscillator, we calibrate the conversion to meters as described in Section \ref{sec:calibration} below. 
Once the signal is linearly dependent on the axial displacement of the particle, it can be used for direct feedback, for which we employ a \textit{Red Pitaya} which filters the homodyne signal around the axial frequency and provides a gain and phase delay to the signal before applying it to a custom-built, bidirectional current amplifier which drives the axial feedback coil.

\begin{figure*}
    \centering
    \includegraphics[width=1\linewidth]{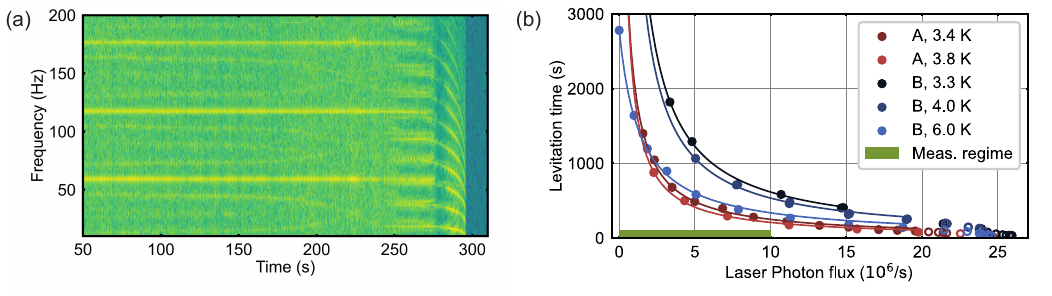}
    \caption{(a) Sample spectrogram of the intensity readout of a levitated particle with a photon flux of around \SI{1.7e7}{\per \second} until the particle quenches and drops. The in-plane radial motion dominates with a frequency of \SI{60}{Hz}. Due to the nonlinear detection caused by the Gaussian reflection profile, higher-order peaks at \SI{120}{Hz} and \SI{180}{Hz} are also visible. (b) Plot of levitation time for two different superconducting particles, A and B, for different laser powers and starting temperatures. The laser power applied is calibrated with the calibration mirror at the start of each measurement. The curves are fitted according to $\tau_\mathrm{lev} = a/(b+n_\mathrm{laser})$.
    The parameter region in which all other measurements in the paper were taken is indicated by the green area.}
    \label{fig:Quench_Time}
\end{figure*}

\subsection{Levitation time}\label{sec:lifetime}
As soon as the particle begins levitating, it loses its direct thermal connection to the \SI{3}{K} bath and begins to heat up. 
By installing a \SI{3}{mm} pinhole in the \SI{30}{K} stage, ambient thermal radiation was reduced to a minimum, allowing for levitation on the order of hours without applied laser power. When laser light is applied at low powers, the particle`s levitation lifetime behaves as
\begin{align}
\tau_\mathrm{lev} = \frac{\Delta E}{P} \;,
\end{align}
where $\Delta E$ is an energy budget for the particle that depends on its starting temperature (and critical temperature),
and $P$ is heat absorbed by the particle.
The power absorbed by the particle will be some fraction of the applied laser power and will depend on its reflectivity (for lead, this is \SI{63}{\%} \cite{Polyanskiy_2024}).
In Fig. \ref{fig:Quench_Time} we show the lifetime dependence on laser power for two different particles, A and B, for different starting temperatures. 

At laser fluxes above \SI{20e6}{photon/s}, we observe a sudden drop in lifetime, the precise details of which are not fully understood.
Furthermore, as seen in Fig. \ref{fig:Quench_Time}, during the final $10$--$20\%$ of flight time, an additional lower-frequency motional mode appears. These modes are not reproducible and are probably related to the rotation or libration of the imperfect particle. During the quench, frequencies then collapse to zero as the superconducting portion of the sphere decreases, reducing the restoring forces in the trap.
Our readout and cooling protocols are carried out over much shorter time periods and with lower laser power, as indicated in Fig.~\ref{fig:Quench_Time}, such that these levitation lifetime effects do not affect the measurements.

We estimate the thermal energy budget of the particle as follows, approximating it as pure lead.
At low temperature, in the superconducting state, lead has a (combined phononic and electronic) heat capacity of $c = 0.0115\,\left(\frac{T}{\mathrm{K}}\right)^3\,\mathrm{J}\,\mathrm{kg}^{-1}\,\mathrm{K}^{-1}$ \cite{Horowitz_1952}.
Furthermore, lead has a superconducting critical temperature of $T_\mathrm{c} = 7.2\,\mathrm{K}$, and a critical field of $H_0 = 6.4\times10^{4}\,\mathrm{A}/\mathrm{m}$ \cite{Chanin_1972}.
In the center of the trap with field gradients of $100\,\mathrm{T}/\mathrm{m}$ and displacements up to \SI{50}{\micro m}, the particle experiences magnetic fields approaching $5000\,\mathrm{A}/\mathrm{m}$.
Using a simple empirical relationship, $H_\mathrm{c} = H_0\,(1-T^2/{T_\mathrm{c}}^2)$ \cite{Tinkham}, the expected quenching temperature occurs around $T = 6.9\,\mathrm{K}$.
For a \SI{6}{\micro g} particle, integrating $\mathrm{d}E=c\,m\,\mathrm{d}T$ between $T_\mathrm{start}=3.5\,\mathrm{K}$ and $T_\mathrm{end}=6.9\,\mathrm{K}$ gives $\Delta E = 3.6\times10^{-8}\,\mathrm{J}$.
The heat conductivity of lead, even while cold and superconducting, is great enough such that, over a timescale of seconds, the temperature of the sphere is effectively uniform.
If an incident photon flux of $10^{7}/\mathrm{s} \sim 3\,\mathrm{pW}$ was completely absorbed by the levitating particle as heat (in reality, it must be far less due to reflection), the corresponding levitation lifetime would be $\tau_\mathrm{lev} \sim 12\,000\,\mathrm{s}$.
This is over an order of magnitude greater than what is observed, suggesting that the quench dynamics is governed by more complex processes.

A photon with a wavelength of \SI{637}{nm} has an energy of \SI{1.946}{eV}, far greater than the superconducting band gap of lead (\SI{2.73}{meV}); thus, direct impact will break Cooper pairs and excite quasiparticles. However, the recombination time is on the scale of nanoseconds \cite{Ginsberg_1962,Kaplan_1976} such that, at low laser photon fluxes, the quasiparticle density equilibrates. At higher photon fluxes, a splitting rate exceeding the recombination rate may explain the abrupt drop in levitation lifetime.
If the coupling between electron temperature and phonon temperature is sufficiently poor (over thousands of seconds), this could explain the lower energy budget. Using only the electron heat capacity of lead, $c = 0.0010\,\left(T/\mathrm{K}\right)^3\,\mathrm{J}\,\mathrm{kg}^{-1}\,\mathrm{K}^{-1}$ \cite{Horowitz_1952}, the expected lifetime from \SI{3}{pW} of heating is only around \SI{1000}{s}.

\begin{figure*}[t!]
    \centering
    \includegraphics[width=1\linewidth]{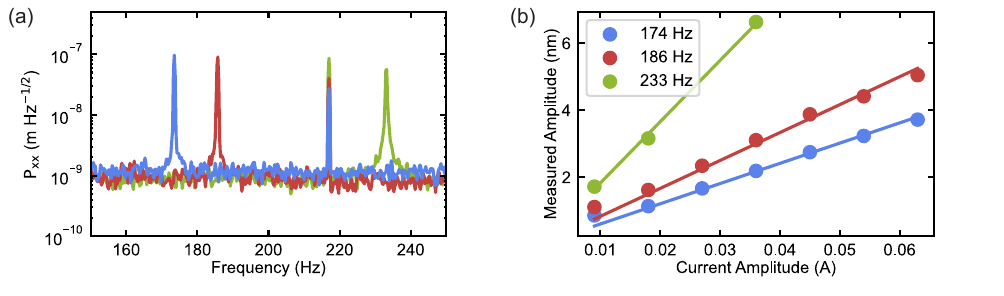}
    \caption{(a) The power spectral densities for the same drive amplitude applied at \SI{ 217}{Hz} as the trap frequency is varied between 175, 185, and \SI{233}{Hz}. (b) The measured amplitudes at drive frequency \SI{217}{Hz} while the trap is detuned to the indicated frequency. The amplitude is calibrated by the known dimensions of the driving solenoid and its distance to the levitated particle.}
    \label{fig:Drive_psd}
\end{figure*}

\subsection{Probe-tone calibration of the feedback}
\label{sec:calibration}
When interferometers are used to probe a relative distance, the conversion from the signal (in most cases, voltage) to meters is determined by the wavelength of the light. In some cases, where effects other than arm-length difference change the phase of the interferometer, it is necessary to calibrate the conversion by other means \cite{Hebestreit_2018}.
In our case, the roughness of the particle in combination with free rotation requires a strong lock to stabilize the phase of the interferometer and avoid leaving the linear regime. Hence, the effective suppression of the signal is not obvious. 
The equilibrium displacement of the particle in the magnetic quadrupole field, with the magnetic field gradient $\mathrm{d}B_\mathrm{trap}/\mathrm{d}z$, due to a constant external magnetic field offset, $B_\mathrm{ext}$, can be approximated by 
\begin{align}
    \Delta z = \frac{B_\mathrm{ext}}{\mathrm{d}B_\mathrm{trap}/\mathrm{d}z} \;.
\end{align} 
The gradient $dB_\mathrm{trap}/\mathrm{d}z$ can be calculated from the frequency of the sphere with the density of the particle given by $\rho = \SI{1.1e4}{kg/m^3}$ \cite{Hofer_2023}. 
From the displacement $\Delta z$, the force acting on the mechanics can be calculated as
\begin{align}
    F = m\,\omega_z^2\,\Delta z \;.
\end{align}
By applying an oscillating current with frequency $\omega_\mathrm{dr}$, we apply an oscillating force $F=F_0\,\sin[\omega_\mathrm{dr}\,t]$, which produces a position modulation of
\begin{align}
    x = \frac{F_0}{m \sqrt2 \left(\omega_z^2-\omega_\mathrm{dr}^2 \right) }\;.
\end{align}
Here we use the assumption that $\gamma^2\,\omega_\mathrm{dr} \ll \left(\omega_0^2-\omega_\mathrm{dr}^2 \right)^2$ where $\gamma$ is the damping rate of the mechanical oscillator \cite{Hebestreit_2018}. As illustrated in Fig.~\ref{fig:Drive_psd}, we measured the response of the oscillator to an external probe tone at different resonance frequencies. At each frequency, we fit the linear relation between the response and the set current amplitude, as illustrated in Fig. \ref{fig:Drive_psd}. By comparing the measured signal with the known properties of the feedback coil, we can reconstruct the noise floor of our detection.

\begin{figure}[b!]
    \centering
    \includegraphics[width=1\linewidth]{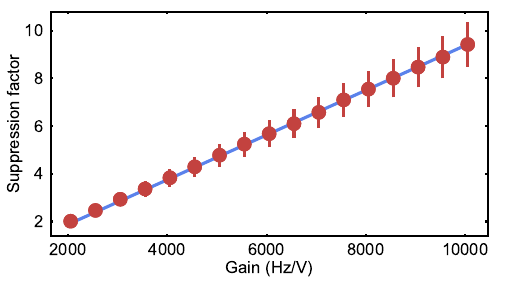}
    \caption{The suppression factor of the lock for different gains is calculated by applying a sinusoidal voltage, with known displacement amplitude, to the piezo-mounted mirror, mimicking the motion of the levitated particle.}
    \label{fig:mirror_cal}
\end{figure}

\subsection{Calibration on a moving mirror}
As a second, independent calibration, the optical path was altered to reflect off a mirror mounted on a piezoelectric crystal, instead of the levitated particle. The motion of this mirror can be precisely controlled by the voltage applied to the crystal. To mimic the motion of the levitated superconductor, the frequency of the mirror displacement was set to \SI{217}{Hz} and the amplitude to $\lambda/{8}\,(\pm10\%)$, where $\lambda=637\,\mathrm{nm}$ is the laser wavelength. To obtain the suppression of the signal due to the lock, we divide the actual amplitude by the measured amplitude and derive a linear relation between the gain and the suppression, as shown in Fig.~\ref{fig:mirror_cal}. The two methods agree within the uncertainty. In the probe tone calibration, we measure a suppression ratio of $7.95\pm1.17$ at \SI{8000}{\hertz /\volt} compared to $7.5\pm 0.75$ for the piezo mirror calibration.

\subsection{Feedback cooling}
A classical harmonic oscillator is governed by the equation of motion,
\begin{align*}
\mass\,\ddot{x}+\mass\,\gamma\,\dot{x}+\mass\,\omg^2\,x=F_\noise+F_\feedback \;,
\end{align*}
where $F_\noise$ is noise, and $F_\feedback$ a feedback drive.
If $x(t)$ can be accurately and continually measured, then a drive $F_\feedback(t)=-\mass\,\gammafb\,\dot{x}(t)$ can be applied to damp the resonator and cool it below its normal (thermal or otherwise) occupancy from $F_\noise$.
Classically, the (two-sided) thermal noise is $S_{F_\thermaln F_\thermaln}=2\,\gamma\,\mass\,\kB\,\temp$, due to the fluctuation-dissipation theorem \cite{Kubo_1966}.

\subsection{Sensitivity requirements}
The achievable damping depends on the noise in the system. 
Including some error in the applied drive, its Fourier transform is $F(\omega) = \ii\,\omega\,\,\mass\,\gammafb\,\bigl(x(\omega)+\dx\bigr)$ where $\dx$ is some error in the applied feedback.
The full Fourier-transformed equation of motion is then
\begin{align*}
x(\omega) = \frac{F_\noise+\ii\,\omega\,\mass\,\gammafb\,\dx}{\mass\,\Bigl( \omg^2-\omega^2-\ii\,\omega\,(\gamma+\gammafb) \Bigr)} \;,
\end{align*}
and taking the (two-sided) spectral density,
\begin{align*}
S_{xx}(\omega)
= \frac{S_{F_\noise F_\noise} +\omega^2\,\mass^2\,\gammafb^2\,S_{\dx \dx}(\omega)}{\mass^2\,\Bigl( (\omg^2-\omega^2)^2 + \omega^2\,(\gamma+\gammafb)^2 \Bigr)} \;,
\end{align*}
which we can integrate to get $\angles{x^2}=\frac{1}{2\,\pi}\int_{-\infty}^{\infty}S_{x x}(\omega)\,\dd\omega$.

For a high-$Q$ resonator, we can approximate
\begin{multline}
\frac{1}{(\omg^2-\omega^2)^2 + \omega^2\,\gamma^2 } \approx \\
\frac{1}{4\,\omg^2}\,\left(\frac{1}{(\omega-\omg)^2+\frac{\gamma^2}{4}} + \frac{1}{(\omega+\omg)^2+\frac{\gamma^2}{4}} \right) \nonumber\;,
\end{multline}
and
\begin{align*}
\frac{f(\omega)}{(\omega-\omg)^2+\frac{\gamma^2}{4}} \approx \frac{f(\omg)}{(\omega-\omg)^2+\frac{\gamma^2}{4}} \;,
\end{align*}
when $f(\omega)$ is sufficiently constant over the resonator bandwidth.
This will allow us to use the integral
\begin{align*}
\int_{-\infty}^{\infty} \frac{1}{(\omega-\omg)^2+\frac{\gamma^2}{4}}\,\dd\omega = \frac{2\,\pi}{\gamma} \;.
\end{align*}
These relationships can be summarized to yield the expectation value,
\begin{align*}
\angles{x^2} = \frac{S_{F_\noise F_\noise}}{2\,\mass^2\,\omg^2\,(\gamma+\gammafb)}+\frac{{\gammafb}^2\,S_{\dx \dx}}{2\,(\gamma+\gammafb)} \;.
\end{align*}
An effective temperature can be stated by equating to the harmonic oscillator energy, $\kB\,\temp_\mathrm{eff} = \mass\,\omg^2\,\angles{x^2}$.

The minimal displacement variance with optimal feedback parameter can be found by $\frac{\dd\,\angles{x^2}}{\dd \gammafb} = 0$, giving
\begin{align*}
\gammafb^\optimal = \sqrt{\frac{S_{F_\noise F_\noise}}{\mass^2\,\omg^2\,S_{\dx \dx}}+\gamma^2} - \gamma
\approx \sqrt{\frac{S_{F_\noise F_\noise}}{\mass^2\,\omg^2\,S_{\dx \dx}}}
\;,
\end{align*}
and minimum variance
\begin{align*}
\angles{x^2}_\mathrm{min}
=&\; S_{\dx \dx}\,\sqrt{\frac{S_{F_\noise F_\noise}}{\mass^2\,\omg^2\,S_{\dx \dx}}+\gamma^2} - S_{\dx \dx}\,\gamma
\approx \sqrt{\frac{S_{F_\noise F_\noise}\,S_{\dx \dx}}{\mass^2\,\omg^2}} \\\\
=&\; (\gamma+\gammafb^\optimal)\, S_{\dx \dx} \approx \gammafb^\optimal\, S_{\dx \dx}
\;,
\end{align*}
where the approximations correspond to high-noise and large feedback damping $\gammafb \gg \gamma$.
Under this approximation, cooling is limited to
\begin{align}
S_{\dx \dx} \, S_{F_\noise F_\noise} \approx \hbar^2\,\left(\bar{n}_\mathrm{min}+\frac{1}{2}\right)^2
\label{eq:SeeTarget}\;,
\end{align}
where we have $\angles{x^2}= \frac{\hbar}{\mass\,\omg}\,\left(\bar{n}+\frac{1}{2}\right)$.

In a continuous feedback system, this error will depend on the measurement noise for $x$. Most generally, our measurement will be $\xmeas=x+\mx$, where $\mx$ is added measurement noise. With no additional noise in the feedback loop, $\dx=\mx$.
Taking care of cross-terms appropriately, the measured spectral density is \cite{Poggio_2007}
\begin{multline}
S_{(x+\mx)(x+\mx)}(\omega)
= \\
\frac{S_{F_\noise F_\noise} +\mass^2\,\Bigl( (\omg^2-\omega^2)^2 +\omega^2\,\gamma^2\Bigr)\,S_{\mx \mx}(\omega)}{\mass^2\,\Bigl( (\omg^2-\omega^2)^2 + \omega^2\,(\gamma+\gammafb)^2 \Bigr)}
\nonumber \;,
\end{multline}
which for constant $S_{\mx \mx}$ is not integrable. For sufficiently large $\gammafb$ this produces a dip below the noise background in what is known as squashing.
Away from the resonance, $S_{(x+\mx)(x+\mx)}(\infty)=S_{\mx \mx}$.
When optimal feedback is applied, $S_{(x+\mx)(x+\mx)}(\omg)=S_{\mx \mx}$.

\subsection{Measurement and back-action noise}
Generally, our measurement imprecision noise will be related to back-action force noise on the oscillator.
The standard quantum limit for this is \cite{Clerk_2010,Aspelmeyer_2014}
\begin{align*}
  S_{x x}^\mathrm{imp}\,S_{F_\backaction F_\backaction} = \frac{\hbar^2}{4} \;.
\end{align*}
Including some detection efficiency, $\eta_{\mathrm{det}}$, and thermal noise,
\begin{align*}
S_{\sigma\sigma} =&\; \frac{S_{x x}^\mathrm{imp}}{\eta_{\mathrm{det}}} \;, &
S_{F_\noise F_\noise} =&\; S_{F_\backaction F_\backaction} + S_{F_\thermaln F_\thermaln} \;.
\end{align*}
Now, defining the back-action decoherence rate and thermal decoherence rate
\begin{align*}
\gammaba=&\;\frac{S_{F_\backaction F_\backaction}}{4\,{p_\mathrm{zpf}}^2} = \frac{S_{F_\backaction F_\backaction}}{2\,\hbar\,\mass\,\omg} \;, &
\gammathermal =&\; \bar{n}_\thermaln(\omg)\,\gamma \;,
\end{align*}
where $\bar{n}_\thermaln(\omg) \approx \frac{\kB\,\temp}{\hbar\,\omg}$,
the quantum cooperativity is \cite{Magrini_2021}
\begin{align*}
\mathcal{C}_\mathrm{q} = \frac{\gammaba}{\gammathermal} = \frac{S_{F_\backaction F_\backaction}}{S_{F_\thermaln F_\thermaln}} \;,
\end{align*}
and ground state cooling requires, from \eqref{eq:SeeTarget},
\begin{align*}
\bar{n} < 1 \quad\Rightarrow\quad
\mathcal{C}_\mathrm{q} > \frac{1}{9\,\eta_\mathrm{det}-1}
\;.
\end{align*}

For photons, with wavenumber $k=\omega/c$, measuring the position of some object, the associated noise spectral densities are \cite{Clerk_2010}
\begin{align*}
S_{x x}^\mathrm{imp} =&\; \frac{1}{16\,k^2\,n_\mathrm{in}} \;, &
S_{F_\backaction F_\backaction} =&\; 4\,\hbar^2\,k^2\,n_\mathrm{in} \;,
\end{align*}
where $n_\mathrm{in}$ is the number of photons per second.

\subsection{Optomechanics}
An optomechanical system \cite{Aspelmeyer_2014,Povey_Thesis} has imprecision noise
\begin{align*}
\bar{S}_{\hat{x} \hat{x}}^\mathrm{imp}(\omg)
=&\;\frac{{\gammaoptic}^2}{16\,G^2\,\noptic\,\gammaopticcoupling} \, \left(1+\frac{4\,\omg^2}{{\gammaoptic}^2}\right) \;,
\end{align*}
and back-action noise
\begin{align*}
S_{F_\backaction F_\backaction}(\omg) = \frac{4\,\hbar^2\,G^2\,\noptic}{\gammaoptic}\,\left(1+\frac{4\,\omg^2}{{\gammaoptic}^2}\right)^{-1} \;.
\end{align*}
Here $G$ is the optical frequency shift per displacement in the cavity.
A one-sided, infinitely over-coupled optical cavity has $\gammaoptic=\gammaopticcoupling$ and $\noptic = (4\,\gammaopticcoupling/\gammaoptic^2) \bar{n}_\mathrm{in} = 4\,\bar{n}_\mathrm{in}/\gammaoptic$.

Force white-noise (such as a thermal or back-action) can always be cooled by increasing the feed-back gain.
Considering only back-action, (effective) thermal, and imprecision noise, with detector efficiency $\eta_\mathrm{det}$, we have
$S_{\dx \dx} = S_{\mx \mx} = \frac{1}{\eta_\mathrm{det}}\,\bar{S}_{\hat{x} \hat{x}}^\mathrm{imp}$ and $S_{F_\noise F_\noise} = S_{F_\backaction F_\backaction}+S_{F_\thermaln F_\thermaln}$.
Combining this detector efficiency with the external coupling ratio, define $\eta=\eta_\mathrm{det}\,\gammaopticcoupling/\gammaoptic$,
then substituting everything into \eqref{eq:SeeTarget},
\begin{align*}
n<1\quad\Rightarrow\quad
\noptic >&\; \frac{\gammaoptic\,S_{F_\thermaln F_\thermaln}}{4\,\hbar^2\,G^2\,\left(9\,\eta-1\right)} \\\\
\Rightarrow\quad \frac{\mathcal{C}_{\optic\mech}}{\bar{n}_\thermaln(\omgmech)} >&\;  \frac{1}{9\,\eta-1}
\quad:\quad \mathcal{C}_{\optic\mech} = \frac{4\,g^2\,n_\mathrm{cav}}{\gammaoptic\,\gammamech}
\;,
\end{align*}
where $\mathcal{C}_{\optic\mech}$ is the optomechanical cooperativity and $\bar{n}_\thermaln(\omgmech)$ is the thermal mechanical population.
The quantum cooperativity is $\mathcal{C}_\mathrm{q}=\mathcal{C}_{\optic\mech}/\bar{n}_\thermaln(\omgmech)$.

In addition to the quantum-limited noise, additional measurement noise $S_{\sigma^+ \sigma^+}$ beyond the shot-noise imprecision needs to be considered. As well as force noise $S_{F_\noise^+ F_\noise^+}$ excluding back-action, which can be described by a total effective temperature.
With only $\bar{S}_{\hat{x} \hat{x}}^\mathrm{imp}$ and $S_{F_\backaction F_\backaction}$ dependent on cavity photon number, there exists an optimal value
\begin{align*}
\underset{\mathrm{opt}}{n_{\mathrm{cav}}} = \frac{\gammaoptic}{8\,G^2\,\hbar\,\sqrt{\eta}}\,\sqrt{\frac{S_{F_\noise^+ F_\noise^+}}{S_{\sigma^+ \sigma^+}}}\,\left(1+\frac{4\,\omg^2}{{\gammaoptic}^2}\right) \;.
\end{align*}
Putting this into \eqref{eq:SeeTarget}, we find ground state cooling requires
\begin{align*}
n<1\quad\Rightarrow\quad
\sqrt{S_{\sigma^+ \sigma^+}\,S_{F_\noise^+ F_\noise^+}} < \frac{\hbar}{2} \, \left(3-\frac{1}{\sqrt{\eta}}\right)
\end{align*}
and if our force noise can be described by an (effective) temperature, $S_{F_\noise^+ F_\noise^+}=2\,\mass\,\gammamech\,\kB\,\temp$, we need
\begin{align*}
n<1\quad\Rightarrow\quad
S_{\sigma^+ \sigma^+} < \frac{1}{\eta}\,\frac{\hbar^2}{8\,\mass\,\gammamech\,\kB\,\temp}\,\left(3\,\sqrt{\eta}-1\right)^2 \;,
\end{align*}
with $\eta>1/9$.

A Fabry-Pérot optical cavity with one moving mirror has optomechanical coupling factor $G\approx-\omega_\mathrm{cav}/L$ and finesse $F=c\,\pi/(\kappa\,L)$,
where $\omega_\mathrm{cav}$ is the angular resonance frequency of interest and $L$ is the cavity length.

\bibliography{bib}

\begin{thebibliography}{63}%
\makeatletter
\providecommand \@ifxundefined [1]{%
 \@ifx{#1\undefined}
}%
\providecommand \@ifnum [1]{%
 \ifnum #1\expandafter \@firstoftwo
 \else \expandafter \@secondoftwo
 \fi
}%
\providecommand \@ifx [1]{%
 \ifx #1\expandafter \@firstoftwo
 \else \expandafter \@secondoftwo
 \fi
}%
\providecommand \natexlab [1]{#1}%
\providecommand \enquote  [1]{``#1''}%
\providecommand \bibnamefont  [1]{#1}%
\providecommand \bibfnamefont [1]{#1}%
\providecommand \citenamefont [1]{#1}%
\providecommand \href@noop [0]{\@secondoftwo}%
\providecommand \href [0]{\begingroup \@sanitize@url \@href}%
\providecommand \@href[1]{\@@startlink{#1}\@@href}%
\providecommand \@@href[1]{\endgroup#1\@@endlink}%
\providecommand \@sanitize@url [0]{\catcode `\\12\catcode `\$12\catcode `\&12\catcode `\#12\catcode `\^12\catcode `\_12\catcode `\%12\relax}%
\providecommand \@@startlink[1]{}%
\providecommand \@@endlink[0]{}%
\providecommand \url  [0]{\begingroup\@sanitize@url \@url }%
\providecommand \@url [1]{\endgroup\@href {#1}{\urlprefix }}%
\providecommand \urlprefix  [0]{URL }%
\providecommand \Eprint [0]{\href }%
\providecommand \doibase [0]{https://doi.org/}%
\providecommand \selectlanguage [0]{\@gobble}%
\providecommand \bibinfo  [0]{\@secondoftwo}%
\providecommand \bibfield  [0]{\@secondoftwo}%
\providecommand \translation [1]{[#1]}%
\providecommand \BibitemOpen [0]{}%
\providecommand \bibitemStop [0]{}%
\providecommand \bibitemNoStop [0]{.\EOS\space}%
\providecommand \EOS [0]{\spacefactor3000\relax}%
\providecommand \BibitemShut  [1]{\csname bibitem#1\endcsname}%
\let\auto@bib@innerbib\@empty
\bibitem [{\citenamefont {O’Connell}\ \emph {et~al.}(2010)\citenamefont {O’Connell}, \citenamefont {Hofheinz}, \citenamefont {Ansmann}, \citenamefont {Bialczak}, \citenamefont {Lenander}, \citenamefont {Lucero}, \citenamefont {Neeley}, \citenamefont {Sank}, \citenamefont {Wang}, \citenamefont {Weides}, \citenamefont {Wenner}, \citenamefont {Martinis},\ and\ \citenamefont {Cleland}}]{Connell_2010}%
  \BibitemOpen
  \bibfield  {author} {\bibinfo {author} {\bibfnamefont {A.}~\bibnamefont {O’Connell}}, \bibinfo {author} {\bibfnamefont {M.}~\bibnamefont {Hofheinz}}, \bibinfo {author} {\bibfnamefont {M.}~\bibnamefont {Ansmann}}, \bibinfo {author} {\bibfnamefont {R.}~\bibnamefont {Bialczak}}, \bibinfo {author} {\bibfnamefont {M.}~\bibnamefont {Lenander}}, \bibinfo {author} {\bibfnamefont {E.}~\bibnamefont {Lucero}}, \bibinfo {author} {\bibfnamefont {M.}~\bibnamefont {Neeley}}, \bibinfo {author} {\bibfnamefont {D.}~\bibnamefont {Sank}}, \bibinfo {author} {\bibfnamefont {H.}~\bibnamefont {Wang}}, \bibinfo {author} {\bibfnamefont {M.}~\bibnamefont {Weides}}, \bibinfo {author} {\bibfnamefont {J.}~\bibnamefont {Wenner}}, \bibinfo {author} {\bibfnamefont {J.}~\bibnamefont {Martinis}},\ and\ \bibinfo {author} {\bibfnamefont {A.}~\bibnamefont {Cleland}},\ }\bibfield  {title} {\bibinfo {title} {Quantum ground state and single-phonon control of a mechanical resonator},\ }\href {https://doi.org/10.1038/nature08967} {\bibfield
  {journal} {\bibinfo  {journal} {Nature}\ }\textbf {\bibinfo {volume} {464}},\ \bibinfo {pages} {697} (\bibinfo {year} {2010})}\BibitemShut {NoStop}%
\bibitem [{\citenamefont {Palomaki}\ \emph {et~al.}(2013)\citenamefont {Palomaki}, \citenamefont {Harlow}, \citenamefont {Teufel}, \citenamefont {Simmonds},\ and\ \citenamefont {Lehnert}}]{Palomaki_2013}%
  \BibitemOpen
  \bibfield  {author} {\bibinfo {author} {\bibfnamefont {T.}~\bibnamefont {Palomaki}}, \bibinfo {author} {\bibfnamefont {J.}~\bibnamefont {Harlow}}, \bibinfo {author} {\bibfnamefont {J.}~\bibnamefont {Teufel}}, \bibinfo {author} {\bibfnamefont {R.}~\bibnamefont {Simmonds}},\ and\ \bibinfo {author} {\bibfnamefont {K.}~\bibnamefont {Lehnert}},\ }\bibfield  {title} {\bibinfo {title} {Coherent state transfer between itinerant microwave fields and a mechanical oscillator},\ }\href {https://doi.org/10.1038/nature11915} {\bibfield  {journal} {\bibinfo  {journal} {Nature}\ }\textbf {\bibinfo {volume} {495}},\ \bibinfo {pages} {210} (\bibinfo {year} {2013})}\BibitemShut {NoStop}%
\bibitem [{\citenamefont {Wollman}\ \emph {et~al.}(2015)\citenamefont {Wollman}, \citenamefont {Lei}, \citenamefont {Weinstein}, \citenamefont {Suh}, \citenamefont {Kronwald}, \citenamefont {Marquardt}, \citenamefont {Clerk},\ and\ \citenamefont {Schwab}}]{Wollman_2015}%
  \BibitemOpen
  \bibfield  {author} {\bibinfo {author} {\bibfnamefont {E.}~\bibnamefont {Wollman}}, \bibinfo {author} {\bibfnamefont {C.~U.}\ \bibnamefont {Lei}}, \bibinfo {author} {\bibfnamefont {A.}~\bibnamefont {Weinstein}}, \bibinfo {author} {\bibfnamefont {J.}~\bibnamefont {Suh}}, \bibinfo {author} {\bibfnamefont {A.}~\bibnamefont {Kronwald}}, \bibinfo {author} {\bibfnamefont {F.}~\bibnamefont {Marquardt}}, \bibinfo {author} {\bibfnamefont {A.}~\bibnamefont {Clerk}},\ and\ \bibinfo {author} {\bibfnamefont {K.}~\bibnamefont {Schwab}},\ }\bibfield  {title} {\bibinfo {title} {Quantum mechanics. quantum squeezing of motion in a mechanical resonator},\ }\href {https://doi.org/10.1126/science.aac5138} {\bibfield  {journal} {\bibinfo  {journal} {Science}\ }\textbf {\bibinfo {volume} {349}} (\bibinfo {year} {2015})}\BibitemShut {NoStop}%
\bibitem [{\citenamefont {Riedinger}\ \emph {et~al.}(2018)\citenamefont {Riedinger}, \citenamefont {Wallucks}, \citenamefont {Marinkovic}, \citenamefont {L{\"o}schnauer}, \citenamefont {Aspelmeyer}, \citenamefont {Hong},\ and\ \citenamefont {Gr{\"o}blacher}}]{Riedinger_2018}%
  \BibitemOpen
  \bibfield  {author} {\bibinfo {author} {\bibfnamefont {R.}~\bibnamefont {Riedinger}}, \bibinfo {author} {\bibfnamefont {A.}~\bibnamefont {Wallucks}}, \bibinfo {author} {\bibfnamefont {I.}~\bibnamefont {Marinkovic}}, \bibinfo {author} {\bibfnamefont {C.}~\bibnamefont {L{\"o}schnauer}}, \bibinfo {author} {\bibfnamefont {M.}~\bibnamefont {Aspelmeyer}}, \bibinfo {author} {\bibfnamefont {S.}~\bibnamefont {Hong}},\ and\ \bibinfo {author} {\bibfnamefont {S.}~\bibnamefont {Gr{\"o}blacher}},\ }\bibfield  {title} {\bibinfo {title} {Remote quantum entanglement between two micromechanical oscillators},\ }\href {https://doi.org/10.1038/s41586-018-0036-z} {\bibfield  {journal} {\bibinfo  {journal} {Nature}\ }\textbf {\bibinfo {volume} {556}},\ \bibinfo {pages} {473} (\bibinfo {year} {2018})}\BibitemShut {NoStop}%
\bibitem [{\citenamefont {Ockeloen-Korppi}\ \emph {et~al.}(2018)\citenamefont {Ockeloen-Korppi}, \citenamefont {Damskägg}, \citenamefont {Pirkkalainen}, \citenamefont {Asjad}, \citenamefont {Clerk}, \citenamefont {Massel}, \citenamefont {Woolley},\ and\ \citenamefont {Sillanpää}}]{Ocheloen-Korppi_2018}%
  \BibitemOpen
  \bibfield  {author} {\bibinfo {author} {\bibfnamefont {C.}~\bibnamefont {Ockeloen-Korppi}}, \bibinfo {author} {\bibfnamefont {E.}~\bibnamefont {Damskägg}}, \bibinfo {author} {\bibfnamefont {J.-M.}\ \bibnamefont {Pirkkalainen}}, \bibinfo {author} {\bibfnamefont {M.}~\bibnamefont {Asjad}}, \bibinfo {author} {\bibfnamefont {A.}~\bibnamefont {Clerk}}, \bibinfo {author} {\bibfnamefont {F.}~\bibnamefont {Massel}}, \bibinfo {author} {\bibfnamefont {M.}~\bibnamefont {Woolley}},\ and\ \bibinfo {author} {\bibfnamefont {M.}~\bibnamefont {Sillanpää}},\ }\bibfield  {title} {\bibinfo {title} {Stabilized entanglement of massive mechanical oscillators},\ }\href {https://doi.org/10.1038/s41586-018-0038-x} {\bibfield  {journal} {\bibinfo  {journal} {Nature}\ }\textbf {\bibinfo {volume} {556}} (\bibinfo {year} {2018})}\BibitemShut {NoStop}%
\bibitem [{\citenamefont {Bild}\ \emph {et~al.}(2023)\citenamefont {Bild}, \citenamefont {Fadel}, \citenamefont {Yang}, \citenamefont {von Lüpke}, \citenamefont {Martin}, \citenamefont {Bruno},\ and\ \citenamefont {Chu}}]{Bild_2023}%
  \BibitemOpen
  \bibfield  {author} {\bibinfo {author} {\bibfnamefont {M.}~\bibnamefont {Bild}}, \bibinfo {author} {\bibfnamefont {M.}~\bibnamefont {Fadel}}, \bibinfo {author} {\bibfnamefont {Y.}~\bibnamefont {Yang}}, \bibinfo {author} {\bibfnamefont {U.}~\bibnamefont {von Lüpke}}, \bibinfo {author} {\bibfnamefont {P.}~\bibnamefont {Martin}}, \bibinfo {author} {\bibfnamefont {A.}~\bibnamefont {Bruno}},\ and\ \bibinfo {author} {\bibfnamefont {Y.}~\bibnamefont {Chu}},\ }\bibfield  {title} {\bibinfo {title} {Schrödinger cat states of a 16-microgram mechanical oscillator},\ }\href {https://doi.org/10.1126/science.adf7553} {\bibfield  {journal} {\bibinfo  {journal} {Science}\ }\textbf {\bibinfo {volume} {380}},\ \bibinfo {pages} {274} (\bibinfo {year} {2023})}\BibitemShut {NoStop}%
\bibitem [{\citenamefont {Galinskiy}\ \emph {et~al.}(2024)\citenamefont {Galinskiy}, \citenamefont {Enzian}, \citenamefont {Parniak},\ and\ \citenamefont {Polzik}}]{Galinskiy_2024}%
  \BibitemOpen
  \bibfield  {author} {\bibinfo {author} {\bibfnamefont {I.}~\bibnamefont {Galinskiy}}, \bibinfo {author} {\bibfnamefont {G.}~\bibnamefont {Enzian}}, \bibinfo {author} {\bibfnamefont {M.}~\bibnamefont {Parniak}},\ and\ \bibinfo {author} {\bibfnamefont {E.~S.}\ \bibnamefont {Polzik}},\ }\bibfield  {title} {\bibinfo {title} {Nonclassical correlations between photons and phonons of center-of-mass motion of a mechanical oscillator},\ }\href {https://doi.org/10.1103/PhysRevLett.133.173605} {\bibfield  {journal} {\bibinfo  {journal} {Phys. Rev. Lett.}\ }\textbf {\bibinfo {volume} {133}},\ \bibinfo {pages} {173605} (\bibinfo {year} {2024})}\BibitemShut {NoStop}%
\bibitem [{\citenamefont {Omahen}\ \emph {et~al.}(2025)\citenamefont {Omahen}, \citenamefont {Storz}, \citenamefont {Bild}, \citenamefont {Scheiwiller}, \citenamefont {Fadel},\ and\ \citenamefont {Chu}}]{Omahen_2025}%
  \BibitemOpen
  \bibfield  {author} {\bibinfo {author} {\bibfnamefont {A.}~\bibnamefont {Omahen}}, \bibinfo {author} {\bibfnamefont {S.}~\bibnamefont {Storz}}, \bibinfo {author} {\bibfnamefont {M.}~\bibnamefont {Bild}}, \bibinfo {author} {\bibfnamefont {D.}~\bibnamefont {Scheiwiller}}, \bibinfo {author} {\bibfnamefont {M.}~\bibnamefont {Fadel}},\ and\ \bibinfo {author} {\bibfnamefont {Y.}~\bibnamefont {Chu}},\ }\href {https://arxiv.org/abs/2507.02653} {\bibinfo {title} {An ultra-cold mechanical quantum sensor for tests of new physics}} (\bibinfo {year} {2025}),\ \Eprint {https://arxiv.org/abs/2507.02653} {arXiv:2507.02653 [quant-ph]} \BibitemShut {NoStop}%
\bibitem [{\citenamefont {Neumeier}\ \emph {et~al.}(2024)\citenamefont {Neumeier}, \citenamefont {Ciampini}, \citenamefont {Romero-Isart}, \citenamefont {Aspelmeyer},\ and\ \citenamefont {Kiesel}}]{Neumeier_2024}%
  \BibitemOpen
  \bibfield  {author} {\bibinfo {author} {\bibfnamefont {L.}~\bibnamefont {Neumeier}}, \bibinfo {author} {\bibfnamefont {M.}~\bibnamefont {Ciampini}}, \bibinfo {author} {\bibfnamefont {O.}~\bibnamefont {Romero-Isart}}, \bibinfo {author} {\bibfnamefont {M.}~\bibnamefont {Aspelmeyer}},\ and\ \bibinfo {author} {\bibfnamefont {N.}~\bibnamefont {Kiesel}},\ }\bibfield  {title} {\bibinfo {title} {Fast quantum interference of a nanoparticle via optical potential control},\ }\href {https://doi.org/10.1073/pnas.2306953121} {\bibfield  {journal} {\bibinfo  {journal} {Proceedings of the National Academy of Sciences of the United States of America}\ }\textbf {\bibinfo {volume} {121}},\ \bibinfo {pages} {e2306953121} (\bibinfo {year} {2024})}\BibitemShut {NoStop}%
\bibitem [{\citenamefont {Roda-Llordes}\ \emph {et~al.}(2024)\citenamefont {Roda-Llordes}, \citenamefont {Riera-Campeny}, \citenamefont {Candoli}, \citenamefont {Grochowski},\ and\ \citenamefont {Romero-Isart}}]{Roda_2024}%
  \BibitemOpen
  \bibfield  {author} {\bibinfo {author} {\bibfnamefont {M.}~\bibnamefont {Roda-Llordes}}, \bibinfo {author} {\bibfnamefont {A.}~\bibnamefont {Riera-Campeny}}, \bibinfo {author} {\bibfnamefont {D.}~\bibnamefont {Candoli}}, \bibinfo {author} {\bibfnamefont {T.}~\bibnamefont {Grochowski}, \bibfnamefont {P.}},\ and\ \bibinfo {author} {\bibfnamefont {O.}~\bibnamefont {Romero-Isart}},\ }\bibfield  {title} {\bibinfo {title} {Macroscopic quantum superpositions via dynamics in a wide double-well potential},\ }\bibfield  {journal} {\bibinfo  {journal} {Physical Review Letters}\ }\textbf {\bibinfo {volume} {132}},\ \href {https://doi.org/10.1103/physrevlett.132.023601} {10.1103/physrevlett.132.023601} (\bibinfo {year} {2024})\BibitemShut {NoStop}%
\bibitem [{\citenamefont {Delić}\ \emph {et~al.}(2020)\citenamefont {Delić}, \citenamefont {Reisenbauer}, \citenamefont {Dare}, \citenamefont {Grass}, \citenamefont {Vuletić}, \citenamefont {Kiesel},\ and\ \citenamefont {Aspelmeyer}}]{Delic_2020}%
  \BibitemOpen
  \bibfield  {author} {\bibinfo {author} {\bibfnamefont {U.}~\bibnamefont {Delić}}, \bibinfo {author} {\bibfnamefont {M.}~\bibnamefont {Reisenbauer}}, \bibinfo {author} {\bibfnamefont {K.}~\bibnamefont {Dare}}, \bibinfo {author} {\bibfnamefont {D.}~\bibnamefont {Grass}}, \bibinfo {author} {\bibfnamefont {V.}~\bibnamefont {Vuletić}}, \bibinfo {author} {\bibfnamefont {N.}~\bibnamefont {Kiesel}},\ and\ \bibinfo {author} {\bibfnamefont {M.}~\bibnamefont {Aspelmeyer}},\ }\bibfield  {title} {\bibinfo {title} {Cooling of a levitated nanoparticle to the motional quantum ground state},\ }\href {https://doi.org/10.1126/science.aba3993} {\bibfield  {journal} {\bibinfo  {journal} {Science}\ }\textbf {\bibinfo {volume} {367}},\ \bibinfo {pages} {892} (\bibinfo {year} {2020})}\BibitemShut {NoStop}%
\bibitem [{\citenamefont {Magrini}\ \emph {et~al.}(2021)\citenamefont {Magrini}, \citenamefont {Rosenzweig}, \citenamefont {Bach}, \citenamefont {Deutschmann-Olek}, \citenamefont {Hofer}, \citenamefont {Hong}, \citenamefont {Kiesel}, \citenamefont {Kugi},\ and\ \citenamefont {Aspelmeyer}}]{Magrini_2021}%
  \BibitemOpen
  \bibfield  {author} {\bibinfo {author} {\bibfnamefont {L.}~\bibnamefont {Magrini}}, \bibinfo {author} {\bibfnamefont {P.}~\bibnamefont {Rosenzweig}}, \bibinfo {author} {\bibfnamefont {C.}~\bibnamefont {Bach}}, \bibinfo {author} {\bibfnamefont {A.}~\bibnamefont {Deutschmann-Olek}}, \bibinfo {author} {\bibfnamefont {S.~G.}\ \bibnamefont {Hofer}}, \bibinfo {author} {\bibfnamefont {S.}~\bibnamefont {Hong}}, \bibinfo {author} {\bibfnamefont {N.}~\bibnamefont {Kiesel}}, \bibinfo {author} {\bibfnamefont {A.}~\bibnamefont {Kugi}},\ and\ \bibinfo {author} {\bibfnamefont {M.}~\bibnamefont {Aspelmeyer}},\ }\bibfield  {title} {\bibinfo {title} {Real-time optimal quantum control of mechanical motion at room temperature},\ }\href {https://doi.org/10.1038/s41586-021-03602-3} {\bibfield  {journal} {\bibinfo  {journal} {Nature}\ }\textbf {\bibinfo {volume} {595}},\ \bibinfo {pages} {373} (\bibinfo {year} {2021})}\BibitemShut {NoStop}%
\bibitem [{\citenamefont {Tebbenjohanns}\ \emph {et~al.}(2021)\citenamefont {Tebbenjohanns}, \citenamefont {Mattana}, \citenamefont {Rossi}, \citenamefont {Frimmer},\ and\ \citenamefont {Novotny}}]{Tebbenjohanns_2021}%
  \BibitemOpen
  \bibfield  {author} {\bibinfo {author} {\bibfnamefont {F.}~\bibnamefont {Tebbenjohanns}}, \bibinfo {author} {\bibfnamefont {M.~L.}\ \bibnamefont {Mattana}}, \bibinfo {author} {\bibfnamefont {M.}~\bibnamefont {Rossi}}, \bibinfo {author} {\bibfnamefont {M.}~\bibnamefont {Frimmer}},\ and\ \bibinfo {author} {\bibfnamefont {L.}~\bibnamefont {Novotny}},\ }\bibfield  {title} {\bibinfo {title} {Quantum control of a nanoparticle optically levitated in cryogenic free space},\ }\href {https://doi.org/10.1038/s41586-021-03617-w} {\bibfield  {journal} {\bibinfo  {journal} {Nature}\ }\textbf {\bibinfo {volume} {595}},\ \bibinfo {pages} {378} (\bibinfo {year} {2021})}\BibitemShut {NoStop}%
\bibitem [{\citenamefont {Ranfagni}\ \emph {et~al.}(2022)\citenamefont {Ranfagni}, \citenamefont {B{\o}rkje}, \citenamefont {Marino},\ and\ \citenamefont {Marin}}]{Ranfagni_2022}%
  \BibitemOpen
  \bibfield  {author} {\bibinfo {author} {\bibfnamefont {A.}~\bibnamefont {Ranfagni}}, \bibinfo {author} {\bibfnamefont {K.}~\bibnamefont {B{\o}rkje}}, \bibinfo {author} {\bibfnamefont {F.}~\bibnamefont {Marino}},\ and\ \bibinfo {author} {\bibfnamefont {F.}~\bibnamefont {Marin}},\ }\bibfield  {title} {\bibinfo {title} {Two-dimensional quantum motion of a levitated nanosphere},\ }\href {https://doi.org/10.1103/PhysRevResearch.4.033051} {\bibfield  {journal} {\bibinfo  {journal} {Physical Review Research}\ }\textbf {\bibinfo {volume} {4}},\ \bibinfo {pages} {033051} (\bibinfo {year} {2022})}\BibitemShut {NoStop}%
\bibitem [{\citenamefont {Kamba}\ \emph {et~al.}(2023)\citenamefont {Kamba}, \citenamefont {Shimizu},\ and\ \citenamefont {Aikawa}}]{Kamba_2023}%
  \BibitemOpen
  \bibfield  {author} {\bibinfo {author} {\bibfnamefont {M.}~\bibnamefont {Kamba}}, \bibinfo {author} {\bibfnamefont {R.}~\bibnamefont {Shimizu}},\ and\ \bibinfo {author} {\bibfnamefont {K.}~\bibnamefont {Aikawa}},\ }\bibfield  {title} {\bibinfo {title} {Nanoscale feedback control of six degrees of freedom of a near-sphere},\ }\bibfield  {journal} {\bibinfo  {journal} {Nature Communications}\ }\textbf {\bibinfo {volume} {14}},\ \href {https://doi.org/10.1038/s41467-023-43745-7} {10.1038/s41467-023-43745-7} (\bibinfo {year} {2023})\BibitemShut {NoStop}%
\bibitem [{\citenamefont {Piotrowski}\ \emph {et~al.}(2023)\citenamefont {Piotrowski}, \citenamefont {Windey}, \citenamefont {Vijayan}, \citenamefont {Gonzalez-Ballestero}, \citenamefont {de~los Ríos~Sommer}, \citenamefont {Meyer}, \citenamefont {Quidant}, \citenamefont {Romero-Isart}, \citenamefont {Reimann},\ and\ \citenamefont {Novotny}}]{Piotrowski_2023}%
  \BibitemOpen
  \bibfield  {author} {\bibinfo {author} {\bibfnamefont {J.}~\bibnamefont {Piotrowski}}, \bibinfo {author} {\bibfnamefont {D.}~\bibnamefont {Windey}}, \bibinfo {author} {\bibfnamefont {J.}~\bibnamefont {Vijayan}}, \bibinfo {author} {\bibfnamefont {C.}~\bibnamefont {Gonzalez-Ballestero}}, \bibinfo {author} {\bibfnamefont {A.}~\bibnamefont {de~los Ríos~Sommer}}, \bibinfo {author} {\bibfnamefont {N.}~\bibnamefont {Meyer}}, \bibinfo {author} {\bibfnamefont {R.}~\bibnamefont {Quidant}}, \bibinfo {author} {\bibfnamefont {O.}~\bibnamefont {Romero-Isart}}, \bibinfo {author} {\bibfnamefont {R.}~\bibnamefont {Reimann}},\ and\ \bibinfo {author} {\bibfnamefont {L.}~\bibnamefont {Novotny}},\ }\bibfield  {title} {\bibinfo {title} {Simultaneous ground-state cooling of two mechanical modes of a levitated nanoparticle},\ }\href {https://doi.org/10.1038/s41567-023-01956-1} {\bibfield  {journal} {\bibinfo  {journal} {Nature Physics}\ }\textbf {\bibinfo {volume} {19}},\ \bibinfo {pages} {1009} (\bibinfo {year}
  {2023})}\BibitemShut {NoStop}%
\bibitem [{\citenamefont {Dania}\ \emph {et~al.}(2024)\citenamefont {Dania}, \citenamefont {Bykov}, \citenamefont {Goschin}, \citenamefont {Teller}, \citenamefont {Kassid},\ and\ \citenamefont {Northup}}]{Dania_2024}%
  \BibitemOpen
  \bibfield  {author} {\bibinfo {author} {\bibfnamefont {L.}~\bibnamefont {Dania}}, \bibinfo {author} {\bibfnamefont {D.~S.}\ \bibnamefont {Bykov}}, \bibinfo {author} {\bibfnamefont {F.}~\bibnamefont {Goschin}}, \bibinfo {author} {\bibfnamefont {M.}~\bibnamefont {Teller}}, \bibinfo {author} {\bibfnamefont {A.}~\bibnamefont {Kassid}},\ and\ \bibinfo {author} {\bibfnamefont {T.~E.}\ \bibnamefont {Northup}},\ }\bibfield  {title} {\bibinfo {title} {Ultrahigh quality factor of a levitated nanomechanical oscillator},\ }\href {https://doi.org/10.1103/PhysRevLett.132.133602} {\bibfield  {journal} {\bibinfo  {journal} {Phys. Rev. Lett.}\ }\textbf {\bibinfo {volume} {132}},\ \bibinfo {pages} {133602} (\bibinfo {year} {2024})}\BibitemShut {NoStop}%
\bibitem [{\citenamefont {Goodkind}(1999)}]{Goodkind_1999}%
  \BibitemOpen
  \bibfield  {author} {\bibinfo {author} {\bibfnamefont {J.~M.}\ \bibnamefont {Goodkind}},\ }\bibfield  {title} {\bibinfo {title} {The superconducting gravimeter},\ }\href {https://doi.org/10.1063/1.1150092} {\bibfield  {journal} {\bibinfo  {journal} {Review of Scientific Instruments}\ }\textbf {\bibinfo {volume} {70}},\ \bibinfo {pages} {4131} (\bibinfo {year} {1999})}\BibitemShut {NoStop}%
\bibitem [{\citenamefont {File}\ and\ \citenamefont {Mills}(1963)}]{File_1963}%
  \BibitemOpen
  \bibfield  {author} {\bibinfo {author} {\bibfnamefont {J.}~\bibnamefont {File}}\ and\ \bibinfo {author} {\bibfnamefont {R.~G.}\ \bibnamefont {Mills}},\ }\bibfield  {title} {\bibinfo {title} {Observation of persistent current in a superconducting solenoid},\ }\href {https://doi.org/10.1103/PhysRevLett.10.93} {\bibfield  {journal} {\bibinfo  {journal} {Phys. Rev. Lett.}\ }\textbf {\bibinfo {volume} {10}},\ \bibinfo {pages} {93} (\bibinfo {year} {1963})}\BibitemShut {NoStop}%
\bibitem [{\citenamefont {Romero-Isart}\ \emph {et~al.}(2012)\citenamefont {Romero-Isart}, \citenamefont {Clemente}, \citenamefont {Navau}, \citenamefont {Sanchez},\ and\ \citenamefont {Cirac}}]{Romero_2012}%
  \BibitemOpen
  \bibfield  {author} {\bibinfo {author} {\bibfnamefont {O.}~\bibnamefont {Romero-Isart}}, \bibinfo {author} {\bibfnamefont {L.}~\bibnamefont {Clemente}}, \bibinfo {author} {\bibfnamefont {C.}~\bibnamefont {Navau}}, \bibinfo {author} {\bibfnamefont {A.}~\bibnamefont {Sanchez}},\ and\ \bibinfo {author} {\bibfnamefont {J.}~\bibnamefont {Cirac}},\ }\bibfield  {title} {\bibinfo {title} {Quantum magnetomechanics with levitating superconducting microspheres},\ }\href {https://doi.org/10.1103/PhysRevLett.109.147205} {\bibfield  {journal} {\bibinfo  {journal} {Physical review letters}\ }\textbf {\bibinfo {volume} {109}},\ \bibinfo {pages} {147205} (\bibinfo {year} {2012})}\BibitemShut {NoStop}%
\bibitem [{\citenamefont {Devlin}\ \emph {et~al.}(2019)\citenamefont {Devlin}, \citenamefont {Wursten}, \citenamefont {Harrington}, \citenamefont {Higuchi}, \citenamefont {Blessing}, \citenamefont {Borchert}, \citenamefont {Erlewein}, \citenamefont {Hansen}, \citenamefont {Morgner}, \citenamefont {Bohman}, \citenamefont {Mooser}, \citenamefont {Smorra}, \citenamefont {Wiesinger}, \citenamefont {Blaum}, \citenamefont {Matsuda}, \citenamefont {Ospelkaus}, \citenamefont {Quint}, \citenamefont {Walz}, \citenamefont {Yamazaki},\ and\ \citenamefont {Ulmer}}]{Devlin_2019}%
  \BibitemOpen
  \bibfield  {author} {\bibinfo {author} {\bibfnamefont {J.~A.}\ \bibnamefont {Devlin}}, \bibinfo {author} {\bibfnamefont {E.}~\bibnamefont {Wursten}}, \bibinfo {author} {\bibfnamefont {J.~A.}\ \bibnamefont {Harrington}}, \bibinfo {author} {\bibfnamefont {T.}~\bibnamefont {Higuchi}}, \bibinfo {author} {\bibfnamefont {P.~E.}\ \bibnamefont {Blessing}}, \bibinfo {author} {\bibfnamefont {M.~J.}\ \bibnamefont {Borchert}}, \bibinfo {author} {\bibfnamefont {S.}~\bibnamefont {Erlewein}}, \bibinfo {author} {\bibfnamefont {J.~J.}\ \bibnamefont {Hansen}}, \bibinfo {author} {\bibfnamefont {J.}~\bibnamefont {Morgner}}, \bibinfo {author} {\bibfnamefont {M.~A.}\ \bibnamefont {Bohman}}, \bibinfo {author} {\bibfnamefont {A.~H.}\ \bibnamefont {Mooser}}, \bibinfo {author} {\bibfnamefont {C.}~\bibnamefont {Smorra}}, \bibinfo {author} {\bibfnamefont {M.}~\bibnamefont {Wiesinger}}, \bibinfo {author} {\bibfnamefont {K.}~\bibnamefont {Blaum}}, \bibinfo {author} {\bibfnamefont {Y.}~\bibnamefont {Matsuda}}, \bibinfo {author}
  {\bibfnamefont {C.}~\bibnamefont {Ospelkaus}}, \bibinfo {author} {\bibfnamefont {W.}~\bibnamefont {Quint}}, \bibinfo {author} {\bibfnamefont {J.}~\bibnamefont {Walz}}, \bibinfo {author} {\bibfnamefont {Y.}~\bibnamefont {Yamazaki}},\ and\ \bibinfo {author} {\bibfnamefont {S.}~\bibnamefont {Ulmer}},\ }\bibfield  {title} {\bibinfo {title} {Superconducting solenoid system with adjustable shielding factor for precision measurements of the properties of the antiproton},\ }\href {https://doi.org/10.1103/PhysRevApplied.12.044012} {\bibfield  {journal} {\bibinfo  {journal} {Phys. Rev. Appl.}\ }\textbf {\bibinfo {volume} {12}},\ \bibinfo {pages} {044012} (\bibinfo {year} {2019})}\BibitemShut {NoStop}%
\bibitem [{\citenamefont {Hofer}\ \emph {et~al.}(2023{\natexlab{a}})\citenamefont {Hofer}, \citenamefont {Gross}, \citenamefont {Higgins}, \citenamefont {Huebl}, \citenamefont {Kieler}, \citenamefont {Kleiner}, \citenamefont {Koelle}, \citenamefont {Schmidt}, \citenamefont {Slater}, \citenamefont {Trupke}, \citenamefont {Uhl}, \citenamefont {Weimann}, \citenamefont {Wieczorek},\ and\ \citenamefont {Aspelmeyer}}]{Hofer_23}%
  \BibitemOpen
  \bibfield  {author} {\bibinfo {author} {\bibfnamefont {J.}~\bibnamefont {Hofer}}, \bibinfo {author} {\bibfnamefont {R.}~\bibnamefont {Gross}}, \bibinfo {author} {\bibfnamefont {G.}~\bibnamefont {Higgins}}, \bibinfo {author} {\bibfnamefont {H.}~\bibnamefont {Huebl}}, \bibinfo {author} {\bibfnamefont {O.~F.}\ \bibnamefont {Kieler}}, \bibinfo {author} {\bibfnamefont {R.}~\bibnamefont {Kleiner}}, \bibinfo {author} {\bibfnamefont {D.}~\bibnamefont {Koelle}}, \bibinfo {author} {\bibfnamefont {P.}~\bibnamefont {Schmidt}}, \bibinfo {author} {\bibfnamefont {J.~A.}\ \bibnamefont {Slater}}, \bibinfo {author} {\bibfnamefont {M.}~\bibnamefont {Trupke}}, \bibinfo {author} {\bibfnamefont {K.}~\bibnamefont {Uhl}}, \bibinfo {author} {\bibfnamefont {T.}~\bibnamefont {Weimann}}, \bibinfo {author} {\bibfnamefont {W.}~\bibnamefont {Wieczorek}},\ and\ \bibinfo {author} {\bibfnamefont {M.}~\bibnamefont {Aspelmeyer}},\ }\bibfield  {title} {\bibinfo {title} {High-$q$ magnetic levitation and control of superconducting microspheres
  at millikelvin temperatures},\ }\href {https://doi.org/10.1103/PhysRevLett.131.043603} {\bibfield  {journal} {\bibinfo  {journal} {Phys. Rev. Lett.}\ }\textbf {\bibinfo {volume} {131}},\ \bibinfo {pages} {043603} (\bibinfo {year} {2023}{\natexlab{a}})}\BibitemShut {NoStop}%
\bibitem [{\citenamefont {van Waarde}(2016)}]{vanWaarde2016}%
  \BibitemOpen
  \bibfield  {author} {\bibinfo {author} {\bibfnamefont {B.}~\bibnamefont {van Waarde}},\ }\emph {\bibinfo {title} {The lead zeppelin: a force sensor without a handle}},\ \href@noop {} {Ph.D. thesis},\ \bibinfo  {school} {Leiden University} (\bibinfo {year} {2016})\BibitemShut {NoStop}%
\bibitem [{\citenamefont {Gutierrez~Latorre}\ \emph {et~al.}(2023)\citenamefont {Gutierrez~Latorre}, \citenamefont {Higgins}, \citenamefont {Paradkar}, \citenamefont {Bauch},\ and\ \citenamefont {Wieczorek}}]{Gutierrez_2023}%
  \BibitemOpen
  \bibfield  {author} {\bibinfo {author} {\bibfnamefont {M.}~\bibnamefont {Gutierrez~Latorre}}, \bibinfo {author} {\bibfnamefont {G.}~\bibnamefont {Higgins}}, \bibinfo {author} {\bibfnamefont {A.}~\bibnamefont {Paradkar}}, \bibinfo {author} {\bibfnamefont {T.}~\bibnamefont {Bauch}},\ and\ \bibinfo {author} {\bibfnamefont {W.}~\bibnamefont {Wieczorek}},\ }\bibfield  {title} {\bibinfo {title} {Superconducting microsphere magnetically levitated in an anharmonic potential with integrated magnetic readout},\ }\href {https://doi.org/10.1103/PhysRevApplied.19.054047} {\bibfield  {journal} {\bibinfo  {journal} {Phys. Rev. Appl.}\ }\textbf {\bibinfo {volume} {19}},\ \bibinfo {pages} {054047} (\bibinfo {year} {2023})}\BibitemShut {NoStop}%
\bibitem [{\citenamefont {Schmidt}\ \emph {et~al.}(2024)\citenamefont {Schmidt}, \citenamefont {Claessen}, \citenamefont {Higgins}, \citenamefont {Hofer}, \citenamefont {Hansen}, \citenamefont {Asenbaum}, \citenamefont {Zemlicka}, \citenamefont {Uhl}, \citenamefont {Kleiner}, \citenamefont {Gross}, \citenamefont {Huebl}, \citenamefont {Trupke},\ and\ \citenamefont {Aspelmeyer}}]{Schmidt_2024}%
  \BibitemOpen
  \bibfield  {author} {\bibinfo {author} {\bibfnamefont {P.}~\bibnamefont {Schmidt}}, \bibinfo {author} {\bibfnamefont {R.}~\bibnamefont {Claessen}}, \bibinfo {author} {\bibfnamefont {G.}~\bibnamefont {Higgins}}, \bibinfo {author} {\bibfnamefont {J.}~\bibnamefont {Hofer}}, \bibinfo {author} {\bibfnamefont {J.~J.}\ \bibnamefont {Hansen}}, \bibinfo {author} {\bibfnamefont {P.}~\bibnamefont {Asenbaum}}, \bibinfo {author} {\bibfnamefont {M.}~\bibnamefont {Zemlicka}}, \bibinfo {author} {\bibfnamefont {K.}~\bibnamefont {Uhl}}, \bibinfo {author} {\bibfnamefont {R.}~\bibnamefont {Kleiner}}, \bibinfo {author} {\bibfnamefont {R.}~\bibnamefont {Gross}}, \bibinfo {author} {\bibfnamefont {H.}~\bibnamefont {Huebl}}, \bibinfo {author} {\bibfnamefont {M.}~\bibnamefont {Trupke}},\ and\ \bibinfo {author} {\bibfnamefont {M.}~\bibnamefont {Aspelmeyer}},\ }\bibfield  {title} {\bibinfo {title} {Remote sensing of a levitated superconductor with a flux-tunable microwave cavity},\ }\href
  {https://doi.org/10.1103/PhysRevApplied.22.014078} {\bibfield  {journal} {\bibinfo  {journal} {Phys. Rev. Appl.}\ }\textbf {\bibinfo {volume} {22}},\ \bibinfo {pages} {014078} (\bibinfo {year} {2024})}\BibitemShut {NoStop}%
\bibitem [{\citenamefont {Abbott}\ \emph {et~al.}(2016)\citenamefont {Abbott}, \citenamefont {Abbott}, \citenamefont {Abbott}, \citenamefont {Abernathy}, \citenamefont {Acernese}, \citenamefont {Ackley}, \citenamefont {Adams}, \citenamefont {Adams}, \citenamefont {Addesso}, \citenamefont {Adhikari}, \citenamefont {Adya}, \citenamefont {Affeldt}, \citenamefont {Agathos}, \citenamefont {Agatsuma}, \citenamefont {Aggarwal}, \citenamefont {Aguiar}, \citenamefont {Aiello}, \citenamefont {Ain}, \citenamefont {Ajith},\ and\ \citenamefont {Allen}}]{Abbott_2016}%
  \BibitemOpen
  \bibfield  {author} {\bibinfo {author} {\bibfnamefont {B.~P.}\ \bibnamefont {Abbott}}, \bibinfo {author} {\bibfnamefont {R.}~\bibnamefont {Abbott}}, \bibinfo {author} {\bibfnamefont {T.~D.}\ \bibnamefont {Abbott}}, \bibinfo {author} {\bibfnamefont {M.~R.}\ \bibnamefont {Abernathy}}, \bibinfo {author} {\bibfnamefont {F.}~\bibnamefont {Acernese}}, \bibinfo {author} {\bibfnamefont {K.}~\bibnamefont {Ackley}}, \bibinfo {author} {\bibfnamefont {C.}~\bibnamefont {Adams}}, \bibinfo {author} {\bibfnamefont {T.}~\bibnamefont {Adams}}, \bibinfo {author} {\bibfnamefont {P.}~\bibnamefont {Addesso}}, \bibinfo {author} {\bibfnamefont {R.~X.}\ \bibnamefont {Adhikari}}, \bibinfo {author} {\bibfnamefont {V.~B.}\ \bibnamefont {Adya}}, \bibinfo {author} {\bibfnamefont {C.}~\bibnamefont {Affeldt}}, \bibinfo {author} {\bibfnamefont {M.}~\bibnamefont {Agathos}}, \bibinfo {author} {\bibfnamefont {K.}~\bibnamefont {Agatsuma}}, \bibinfo {author} {\bibfnamefont {N.}~\bibnamefont {Aggarwal}}, \bibinfo {author} {\bibfnamefont {O.~D.}\
  \bibnamefont {Aguiar}}, \bibinfo {author} {\bibfnamefont {L.}~\bibnamefont {Aiello}}, \bibinfo {author} {\bibfnamefont {A.}~\bibnamefont {Ain}}, \bibinfo {author} {\bibfnamefont {P.}~\bibnamefont {Ajith}},\ and\ \bibinfo {author} {\bibfnamefont {B.}~\bibnamefont {Allen}} (\bibinfo {collaboration} {LIGO Scientific Collaboration and Virgo Collaboration}),\ }\bibfield  {title} {\bibinfo {title} {Gw150914: The advanced ligo detectors in the era of first discoveries},\ }\href {https://doi.org/10.1103/PhysRevLett.116.131103} {\bibfield  {journal} {\bibinfo  {journal} {Phys. Rev. Lett.}\ }\textbf {\bibinfo {volume} {116}},\ \bibinfo {pages} {131103} (\bibinfo {year} {2016})}\BibitemShut {NoStop}%
\bibitem [{\citenamefont {Delaney}\ \emph {et~al.}(2022)\citenamefont {Delaney}, \citenamefont {Urmey}, \citenamefont {Mittal}, \citenamefont {Brubaker}, \citenamefont {Kindem}, \citenamefont {Burns}, \citenamefont {Regal},\ and\ \citenamefont {Lehnert}}]{Delaney_2022}%
  \BibitemOpen
  \bibfield  {author} {\bibinfo {author} {\bibfnamefont {R.~D.}\ \bibnamefont {Delaney}}, \bibinfo {author} {\bibfnamefont {M.~D.}\ \bibnamefont {Urmey}}, \bibinfo {author} {\bibfnamefont {S.}~\bibnamefont {Mittal}}, \bibinfo {author} {\bibfnamefont {B.~M.}\ \bibnamefont {Brubaker}}, \bibinfo {author} {\bibfnamefont {J.~M.}\ \bibnamefont {Kindem}}, \bibinfo {author} {\bibfnamefont {P.~S.}\ \bibnamefont {Burns}}, \bibinfo {author} {\bibfnamefont {C.~A.}\ \bibnamefont {Regal}},\ and\ \bibinfo {author} {\bibfnamefont {K.~W.}\ \bibnamefont {Lehnert}},\ }\bibfield  {title} {\bibinfo {title} {Superconducting-qubit readout via low-backaction electro-optic transduction},\ }\href {https://doi.org/10.1038/s41586-022-04720-2} {\bibfield  {journal} {\bibinfo  {journal} {Nature}\ }\textbf {\bibinfo {volume} {606}},\ \bibinfo {pages} {489} (\bibinfo {year} {2022})}\BibitemShut {NoStop}%
\bibitem [{\citenamefont {Meesala}\ \emph {et~al.}(2024)\citenamefont {Meesala}, \citenamefont {Lake}, \citenamefont {Wood}, \citenamefont {Chiappina}, \citenamefont {Zhong}, \citenamefont {Beyer}, \citenamefont {Shaw}, \citenamefont {Jiang},\ and\ \citenamefont {Painter}}]{Meesala_2024}%
  \BibitemOpen
  \bibfield  {author} {\bibinfo {author} {\bibfnamefont {S.}~\bibnamefont {Meesala}}, \bibinfo {author} {\bibfnamefont {D.}~\bibnamefont {Lake}}, \bibinfo {author} {\bibfnamefont {S.}~\bibnamefont {Wood}}, \bibinfo {author} {\bibfnamefont {P.}~\bibnamefont {Chiappina}}, \bibinfo {author} {\bibfnamefont {C.}~\bibnamefont {Zhong}}, \bibinfo {author} {\bibfnamefont {A.~D.}\ \bibnamefont {Beyer}}, \bibinfo {author} {\bibfnamefont {M.~D.}\ \bibnamefont {Shaw}}, \bibinfo {author} {\bibfnamefont {L.}~\bibnamefont {Jiang}},\ and\ \bibinfo {author} {\bibfnamefont {O.}~\bibnamefont {Painter}},\ }\bibfield  {title} {\bibinfo {title} {Quantum entanglement between optical and microwave photonic qubits},\ }\href {https://doi.org/10.1103/PhysRevX.14.031055} {\bibfield  {journal} {\bibinfo  {journal} {Phys. Rev. X}\ }\textbf {\bibinfo {volume} {14}},\ \bibinfo {pages} {031055} (\bibinfo {year} {2024})}\BibitemShut {NoStop}%
\bibitem [{\citenamefont {Arnold}\ \emph {et~al.}(2025)\citenamefont {Arnold}, \citenamefont {Werner}, \citenamefont {Sahu}, \citenamefont {Kapoor}, \citenamefont {Qiu},\ and\ \citenamefont {Fink}}]{Arnold_2025}%
  \BibitemOpen
  \bibfield  {author} {\bibinfo {author} {\bibfnamefont {G.}~\bibnamefont {Arnold}}, \bibinfo {author} {\bibfnamefont {T.}~\bibnamefont {Werner}}, \bibinfo {author} {\bibfnamefont {R.}~\bibnamefont {Sahu}}, \bibinfo {author} {\bibfnamefont {L.~N.}\ \bibnamefont {Kapoor}}, \bibinfo {author} {\bibfnamefont {L.}~\bibnamefont {Qiu}},\ and\ \bibinfo {author} {\bibfnamefont {J.~M.}\ \bibnamefont {Fink}},\ }\bibfield  {title} {\bibinfo {title} {All-optical superconducting qubit readout},\ }\href@noop {} {\bibfield  {journal} {\bibinfo  {journal} {Nature Physics}\ ,\ \bibinfo {pages} {1}} (\bibinfo {year} {2025})}\BibitemShut {NoStop}%
\bibitem [{\citenamefont {Hofer}\ and\ \citenamefont {Aspelmeyer}(2019)}]{Hofer_2019}%
  \BibitemOpen
  \bibfield  {author} {\bibinfo {author} {\bibfnamefont {J.}~\bibnamefont {Hofer}}\ and\ \bibinfo {author} {\bibfnamefont {M.}~\bibnamefont {Aspelmeyer}},\ }\bibfield  {title} {\bibinfo {title} {Analytic solutions to the maxwell–london equations and levitation force for a superconducting sphere in a quadrupole field},\ }\href {https://doi.org/10.1088/1402-4896/ab0c44} {\bibfield  {journal} {\bibinfo  {journal} {Physica Scripta}\ }\textbf {\bibinfo {volume} {94}},\ \bibinfo {pages} {125508} (\bibinfo {year} {2019})}\BibitemShut {NoStop}%
\bibitem [{Sup()}]{Suppl}%
  \BibitemOpen
  \bibfield  {title} {\bibinfo {title} {Supplementary material},\ }\href@noop {} {\ }\BibitemShut {NoStop}%
\bibitem [{\citenamefont {{a}o}\ \emph {et~al.}(2019)\citenamefont {{a}o}, \citenamefont {Martin}, \citenamefont {{a}o M.~S.~Sakamoto}, \citenamefont {Teixeira},\ and\ \citenamefont {Kitano}}]{Luiz_2019}%
  \BibitemOpen
  \bibfield  {author} {\bibinfo {author} {\bibfnamefont {L.~H. V.~F.}\ \bibnamefont {{a}o}}, \bibinfo {author} {\bibfnamefont {R.~I.}\ \bibnamefont {Martin}}, \bibinfo {author} {\bibfnamefont {J.}~\bibnamefont {{a}o M.~S.~Sakamoto}}, \bibinfo {author} {\bibfnamefont {M.~C.~M.}\ \bibnamefont {Teixeira}},\ and\ \bibinfo {author} {\bibfnamefont {C.}~\bibnamefont {Kitano}},\ }\bibfield  {title} {\bibinfo {title} {Wide dynamic range quadrature interferometer with high-gain approach and sliding mode control},\ }\href {https://doi.org/10.1364/OE.27.025031} {\bibfield  {journal} {\bibinfo  {journal} {Opt. Express}\ }\textbf {\bibinfo {volume} {27}},\ \bibinfo {pages} {25031} (\bibinfo {year} {2019})}\BibitemShut {NoStop}%
\bibitem [{\citenamefont {Fisher}\ and\ \citenamefont {Warde}(1979)}]{Fisher_1979}%
  \BibitemOpen
  \bibfield  {author} {\bibinfo {author} {\bibfnamefont {A.~D.}\ \bibnamefont {Fisher}}\ and\ \bibinfo {author} {\bibfnamefont {C.}~\bibnamefont {Warde}},\ }\bibfield  {title} {\bibinfo {title} {Simple closed-loop system for real-time optical phase measurement},\ }\href {https://doi.org/10.1364/OL.4.000131} {\bibfield  {journal} {\bibinfo  {journal} {Opt. Lett.}\ }\textbf {\bibinfo {volume} {4}},\ \bibinfo {pages} {131} (\bibinfo {year} {1979})}\BibitemShut {NoStop}%
\bibitem [{\citenamefont {Clerk}\ \emph {et~al.}(2010)\citenamefont {Clerk}, \citenamefont {Devoret}, \citenamefont {Girvin}, \citenamefont {Marquardt},\ and\ \citenamefont {Schoelkopf}}]{Clerk_2010}%
  \BibitemOpen
  \bibfield  {author} {\bibinfo {author} {\bibfnamefont {A.~A.}\ \bibnamefont {Clerk}}, \bibinfo {author} {\bibfnamefont {M.~H.}\ \bibnamefont {Devoret}}, \bibinfo {author} {\bibfnamefont {S.~M.}\ \bibnamefont {Girvin}}, \bibinfo {author} {\bibfnamefont {F.}~\bibnamefont {Marquardt}},\ and\ \bibinfo {author} {\bibfnamefont {R.~J.}\ \bibnamefont {Schoelkopf}},\ }\bibfield  {title} {\bibinfo {title} {Introduction to quantum noise, measurement, and amplification},\ }\href {https://doi.org/10.1103/RevModPhys.82.1155} {\bibfield  {journal} {\bibinfo  {journal} {Rev. Mod. Phys.}\ }\textbf {\bibinfo {volume} {82}},\ \bibinfo {pages} {1155} (\bibinfo {year} {2010})}\BibitemShut {NoStop}%
\bibitem [{\citenamefont {Poggio}\ \emph {et~al.}(2007)\citenamefont {Poggio}, \citenamefont {Degen}, \citenamefont {Mamin},\ and\ \citenamefont {Rugar}}]{Poggio_2007}%
  \BibitemOpen
  \bibfield  {author} {\bibinfo {author} {\bibfnamefont {M.}~\bibnamefont {Poggio}}, \bibinfo {author} {\bibfnamefont {C.~L.}\ \bibnamefont {Degen}}, \bibinfo {author} {\bibfnamefont {H.~J.}\ \bibnamefont {Mamin}},\ and\ \bibinfo {author} {\bibfnamefont {D.}~\bibnamefont {Rugar}},\ }\bibfield  {title} {\bibinfo {title} {Feedback cooling of a cantilever's fundamental mode below 5 mk},\ }\href {https://doi.org/10.1103/PhysRevLett.99.017201} {\bibfield  {journal} {\bibinfo  {journal} {Phys. Rev. Lett.}\ }\textbf {\bibinfo {volume} {99}},\ \bibinfo {pages} {017201} (\bibinfo {year} {2007})}\BibitemShut {NoStop}%
\bibitem [{\citenamefont {Aspelmeyer}\ \emph {et~al.}(2014)\citenamefont {Aspelmeyer}, \citenamefont {Kippenberg},\ and\ \citenamefont {Marquardt}}]{Aspelmeyer_2014}%
  \BibitemOpen
  \bibfield  {author} {\bibinfo {author} {\bibfnamefont {M.}~\bibnamefont {Aspelmeyer}}, \bibinfo {author} {\bibfnamefont {T.~J.}\ \bibnamefont {Kippenberg}},\ and\ \bibinfo {author} {\bibfnamefont {F.}~\bibnamefont {Marquardt}},\ }\bibfield  {title} {\bibinfo {title} {Cavity optomechanics},\ }\href {https://doi.org/10.1103/RevModPhys.86.1391} {\bibfield  {journal} {\bibinfo  {journal} {Rev. Mod. Phys.}\ }\textbf {\bibinfo {volume} {86}},\ \bibinfo {pages} {1391} (\bibinfo {year} {2014})}\BibitemShut {NoStop}%
\bibitem [{\citenamefont {Ernzer}\ \emph {et~al.}(2023)\citenamefont {Ernzer}, \citenamefont {Bosch~Aguilera}, \citenamefont {Brunelli}, \citenamefont {Schmid}, \citenamefont {Karg}, \citenamefont {Bruder}, \citenamefont {Potts},\ and\ \citenamefont {Treutlein}}]{Ernzer_2023}%
  \BibitemOpen
  \bibfield  {author} {\bibinfo {author} {\bibfnamefont {M.}~\bibnamefont {Ernzer}}, \bibinfo {author} {\bibfnamefont {M.}~\bibnamefont {Bosch~Aguilera}}, \bibinfo {author} {\bibfnamefont {M.}~\bibnamefont {Brunelli}}, \bibinfo {author} {\bibfnamefont {G.-L.}\ \bibnamefont {Schmid}}, \bibinfo {author} {\bibfnamefont {T.~M.}\ \bibnamefont {Karg}}, \bibinfo {author} {\bibfnamefont {C.}~\bibnamefont {Bruder}}, \bibinfo {author} {\bibfnamefont {P.~P.}\ \bibnamefont {Potts}},\ and\ \bibinfo {author} {\bibfnamefont {P.}~\bibnamefont {Treutlein}},\ }\bibfield  {title} {\bibinfo {title} {Optical coherent feedback control of a mechanical oscillator},\ }\href {https://doi.org/10.1103/PhysRevX.13.021023} {\bibfield  {journal} {\bibinfo  {journal} {Phys. Rev. X}\ }\textbf {\bibinfo {volume} {13}},\ \bibinfo {pages} {021023} (\bibinfo {year} {2023})}\BibitemShut {NoStop}%
\bibitem [{\citenamefont {Melo}\ \emph {et~al.}(2025)\citenamefont {Melo}, \citenamefont {Veldhuizen}, \citenamefont {Tomassi}, \citenamefont {Meyer},\ and\ \citenamefont {Quidant}}]{Melo_2025}%
  \BibitemOpen
  \bibfield  {author} {\bibinfo {author} {\bibfnamefont {B.}~\bibnamefont {Melo}}, \bibinfo {author} {\bibfnamefont {D.}~\bibnamefont {Veldhuizen}}, \bibinfo {author} {\bibfnamefont {G.~F.~M.}\ \bibnamefont {Tomassi}}, \bibinfo {author} {\bibfnamefont {N.}~\bibnamefont {Meyer}},\ and\ \bibinfo {author} {\bibfnamefont {R.}~\bibnamefont {Quidant}},\ }\href {https://arxiv.org/abs/2506.21341} {\bibinfo {title} {Cooling of an optically levitated nanoparticle via measurement-free coherent feedback}} (\bibinfo {year} {2025}),\ \Eprint {https://arxiv.org/abs/2506.21341} {arXiv:2506.21341 [quant-ph]} \BibitemShut {NoStop}%
\bibitem [{\citenamefont {Hunger}\ \emph {et~al.}(2010)\citenamefont {Hunger}, \citenamefont {Steinmetz}, \citenamefont {Colombe}, \citenamefont {Deutsch}, \citenamefont {Haensch},\ and\ \citenamefont {Reichel}}]{Hunger_2010}%
  \BibitemOpen
  \bibfield  {author} {\bibinfo {author} {\bibfnamefont {D.}~\bibnamefont {Hunger}}, \bibinfo {author} {\bibfnamefont {T.}~\bibnamefont {Steinmetz}}, \bibinfo {author} {\bibfnamefont {Y.}~\bibnamefont {Colombe}}, \bibinfo {author} {\bibfnamefont {C.}~\bibnamefont {Deutsch}}, \bibinfo {author} {\bibfnamefont {T.}~\bibnamefont {Haensch}},\ and\ \bibinfo {author} {\bibfnamefont {J.}~\bibnamefont {Reichel}},\ }\bibfield  {title} {\bibinfo {title} {Fiber fabry-perot cavity with high finesse},\ }\href {https://doi.org/10.1088/1367-2630/12/6/065038} {\bibfield  {journal} {\bibinfo  {journal} {New J. Phys.}\ }\textbf {\bibinfo {volume} {12}} (\bibinfo {year} {2010})}\BibitemShut {NoStop}%
\bibitem [{\citenamefont {Muller}\ \emph {et~al.}(2010)\citenamefont {Muller}, \citenamefont {Flagg}, \citenamefont {Lawall},\ and\ \citenamefont {Solomon}}]{Muller_2010}%
  \BibitemOpen
  \bibfield  {author} {\bibinfo {author} {\bibfnamefont {A.}~\bibnamefont {Muller}}, \bibinfo {author} {\bibfnamefont {E.~B.}\ \bibnamefont {Flagg}}, \bibinfo {author} {\bibfnamefont {J.~R.}\ \bibnamefont {Lawall}},\ and\ \bibinfo {author} {\bibfnamefont {G.~S.}\ \bibnamefont {Solomon}},\ }\bibfield  {title} {\bibinfo {title} {Ultrahigh-finesse, low-mode-volume fabry-perot microcavity},\ }\href {https://doi.org/10.1364/OL.35.002293} {\bibfield  {journal} {\bibinfo  {journal} {Opt. Lett.}\ }\textbf {\bibinfo {volume} {35}},\ \bibinfo {pages} {2293} (\bibinfo {year} {2010})}\BibitemShut {NoStop}%
\bibitem [{\citenamefont {Fait}\ \emph {et~al.}(2021)\citenamefont {Fait}, \citenamefont {Putz}, \citenamefont {Wachter}, \citenamefont {Schalko}, \citenamefont {Schmid}, \citenamefont {Arndt},\ and\ \citenamefont {Trupke}}]{Fait_2021}%
  \BibitemOpen
  \bibfield  {author} {\bibinfo {author} {\bibfnamefont {J.}~\bibnamefont {Fait}}, \bibinfo {author} {\bibfnamefont {S.}~\bibnamefont {Putz}}, \bibinfo {author} {\bibfnamefont {G.}~\bibnamefont {Wachter}}, \bibinfo {author} {\bibfnamefont {J.}~\bibnamefont {Schalko}}, \bibinfo {author} {\bibfnamefont {U.}~\bibnamefont {Schmid}}, \bibinfo {author} {\bibfnamefont {M.}~\bibnamefont {Arndt}},\ and\ \bibinfo {author} {\bibfnamefont {M.}~\bibnamefont {Trupke}},\ }\bibfield  {title} {\bibinfo {title} {High finesse microcavities in the optical telecom o-band},\ }\href {https://doi.org/10.1063/5.0066620} {\bibfield  {journal} {\bibinfo  {journal} {Applied Physics Letters}\ }\textbf {\bibinfo {volume} {119}},\ \bibinfo {pages} {221112} (\bibinfo {year} {2021})}\BibitemShut {NoStop}%
\bibitem [{\citenamefont {Abdelatief}\ \emph {et~al.}(2025)\citenamefont {Abdelatief}, \citenamefont {Renders}, \citenamefont {Alqedra}, \citenamefont {Hansen}, \citenamefont {Hunger}, \citenamefont {Rippe}, \citenamefont {Walther} \emph {et~al.}}]{Abdelatief_2025}%
  \BibitemOpen
  \bibfield  {author} {\bibinfo {author} {\bibfnamefont {A.~S.}\ \bibnamefont {Abdelatief}}, \bibinfo {author} {\bibfnamefont {A.}~\bibnamefont {Renders}}, \bibinfo {author} {\bibfnamefont {M.~K.}\ \bibnamefont {Alqedra}}, \bibinfo {author} {\bibfnamefont {J.~J.}\ \bibnamefont {Hansen}}, \bibinfo {author} {\bibfnamefont {D.}~\bibnamefont {Hunger}}, \bibinfo {author} {\bibfnamefont {L.}~\bibnamefont {Rippe}}, \bibinfo {author} {\bibfnamefont {A.}~\bibnamefont {Walther}}, \emph {et~al.},\ }\bibfield  {title} {\bibinfo {title} {Micro-cavity length stabilization for fluorescence enhancement using schemes based on higher-order spatial modes},\ }\href@noop {} {\bibfield  {journal} {\bibinfo  {journal} {Review of Scientific Instruments}\ }\textbf {\bibinfo {volume} {96}} (\bibinfo {year} {2025})}\BibitemShut {NoStop}%
\bibitem [{\citenamefont {Gutierrez}\ \emph {et~al.}(2020)\citenamefont {Gutierrez}, \citenamefont {Hofer}, \citenamefont {Rudolph},\ and\ \citenamefont {Wieczorek}}]{Gutierrez_2020}%
  \BibitemOpen
  \bibfield  {author} {\bibinfo {author} {\bibfnamefont {M.}~\bibnamefont {Gutierrez}}, \bibinfo {author} {\bibfnamefont {J.}~\bibnamefont {Hofer}}, \bibinfo {author} {\bibfnamefont {M.}~\bibnamefont {Rudolph}},\ and\ \bibinfo {author} {\bibfnamefont {W.}~\bibnamefont {Wieczorek}},\ }\bibfield  {title} {\bibinfo {title} {Chip-based superconducting traps for levitation of micrometer-sized superconducting particles},\ }\href {https://doi.org/10.1088/1361-6668/aba6e1} {\bibfield  {journal} {\bibinfo  {journal} {Superconductor Science and Technology}\ }\textbf {\bibinfo {volume} {33}},\ \bibinfo {pages} {105002} (\bibinfo {year} {2020})}\BibitemShut {NoStop}%
\bibitem [{\citenamefont {Navau}\ \emph {et~al.}(2021)\citenamefont {Navau}, \citenamefont {Minniberger}, \citenamefont {Trupke},\ and\ \citenamefont {Sanchez}}]{Navau_2021}%
  \BibitemOpen
  \bibfield  {author} {\bibinfo {author} {\bibfnamefont {C.}~\bibnamefont {Navau}}, \bibinfo {author} {\bibfnamefont {S.}~\bibnamefont {Minniberger}}, \bibinfo {author} {\bibfnamefont {M.}~\bibnamefont {Trupke}},\ and\ \bibinfo {author} {\bibfnamefont {A.}~\bibnamefont {Sanchez}},\ }\bibfield  {title} {\bibinfo {title} {Levitation of superconducting microrings for quantum magnetomechanics},\ }\href {https://doi.org/10.1103/PhysRevB.103.174436} {\bibfield  {journal} {\bibinfo  {journal} {Phys. Rev. B}\ }\textbf {\bibinfo {volume} {103}},\ \bibinfo {pages} {174436} (\bibinfo {year} {2021})}\BibitemShut {NoStop}%
\bibitem [{\citenamefont {Hofer}(2024)}]{Hofer_2024}%
  \BibitemOpen
  \bibfield  {author} {\bibinfo {author} {\bibfnamefont {J.}~\bibnamefont {Hofer}},\ }\bibfield  {title} {\bibinfo {title} {A numerical approach to levitated superconductors and its application to a superconducting cylinder in a quadrupole field},\ }\href {https://doi.org/10.1088/1402-4896/ad6bcf} {\bibfield  {journal} {\bibinfo  {journal} {Physica Scripta}\ }\textbf {\bibinfo {volume} {99}},\ \bibinfo {pages} {095526} (\bibinfo {year} {2024})}\BibitemShut {NoStop}%
\bibitem [{\citenamefont {Bort-Soldevila}\ \emph {et~al.}(2024)\citenamefont {Bort-Soldevila}, \citenamefont {Cunill-Subiranas}, \citenamefont {Del-Valle}, \citenamefont {Wieczorek}, \citenamefont {Higgins}, \citenamefont {Trupke},\ and\ \citenamefont {Navau}}]{Bort_2024}%
  \BibitemOpen
  \bibfield  {author} {\bibinfo {author} {\bibfnamefont {N.}~\bibnamefont {Bort-Soldevila}}, \bibinfo {author} {\bibfnamefont {J.}~\bibnamefont {Cunill-Subiranas}}, \bibinfo {author} {\bibfnamefont {N.}~\bibnamefont {Del-Valle}}, \bibinfo {author} {\bibfnamefont {W.}~\bibnamefont {Wieczorek}}, \bibinfo {author} {\bibfnamefont {G.}~\bibnamefont {Higgins}}, \bibinfo {author} {\bibfnamefont {M.}~\bibnamefont {Trupke}},\ and\ \bibinfo {author} {\bibfnamefont {C.}~\bibnamefont {Navau}},\ }\bibfield  {title} {\bibinfo {title} {Modeling magnetically levitated superconducting ellipsoids, cylinders, and cuboids for quantum magnetomechanics},\ }\href {https://doi.org/10.1103/PhysRevResearch.6.043046} {\bibfield  {journal} {\bibinfo  {journal} {Phys. Rev. Res.}\ }\textbf {\bibinfo {volume} {6}},\ \bibinfo {pages} {043046} (\bibinfo {year} {2024})}\BibitemShut {NoStop}%
\bibitem [{\citenamefont {DeWitt}\ and\ \citenamefont {Rickles}(2011)}]{DeWittRickles_2011}%
  \BibitemOpen
  \bibinfo {editor} {\bibfnamefont {C.~M.}\ \bibnamefont {DeWitt}}\ and\ \bibinfo {editor} {\bibfnamefont {D.}~\bibnamefont {Rickles}},\ eds.,\ \href@noop {} {\emph {\bibinfo {title} {The Role of Gravitation in Physics: Report from the 1957 Chapel Hill Conference}}}\ (\bibinfo  {publisher} {Max Planck Research Library for the History and Development of Knowledge},\ \bibinfo {year} {2011})\BibitemShut {NoStop}%
\bibitem [{\citenamefont {Pino}\ \emph {et~al.}(2018)\citenamefont {Pino}, \citenamefont {Prat-Camps}, \citenamefont {Sinha}, \citenamefont {Venkatesh},\ and\ \citenamefont {Romero-Isart}}]{Pino_2018}%
  \BibitemOpen
  \bibfield  {author} {\bibinfo {author} {\bibfnamefont {H.}~\bibnamefont {Pino}}, \bibinfo {author} {\bibfnamefont {J.}~\bibnamefont {Prat-Camps}}, \bibinfo {author} {\bibfnamefont {K.}~\bibnamefont {Sinha}}, \bibinfo {author} {\bibfnamefont {B.~P.}\ \bibnamefont {Venkatesh}},\ and\ \bibinfo {author} {\bibfnamefont {O.}~\bibnamefont {Romero-Isart}},\ }\bibfield  {title} {\bibinfo {title} {On-chip quantum interference of a superconducting microsphere},\ }\href {https://doi.org/10.1088/2058-9565/aa9d15} {\bibfield  {journal} {\bibinfo  {journal} {Quantum Science and Technology}\ }\textbf {\bibinfo {volume} {3}},\ \bibinfo {pages} {025001} (\bibinfo {year} {2018})}\BibitemShut {NoStop}%
\bibitem [{\citenamefont {Krisnanda}\ \emph {et~al.}(2020)\citenamefont {Krisnanda}, \citenamefont {Tham}, \citenamefont {Paternostro},\ and\ \citenamefont {Paterek}}]{Krisnanda_2020}%
  \BibitemOpen
  \bibfield  {author} {\bibinfo {author} {\bibfnamefont {T.}~\bibnamefont {Krisnanda}}, \bibinfo {author} {\bibfnamefont {G.~Y.}\ \bibnamefont {Tham}}, \bibinfo {author} {\bibfnamefont {M.}~\bibnamefont {Paternostro}},\ and\ \bibinfo {author} {\bibfnamefont {T.}~\bibnamefont {Paterek}},\ }\bibfield  {title} {\bibinfo {title} {Observable quantum entanglement due to gravity},\ }\href@noop {} {\bibfield  {journal} {\bibinfo  {journal} {npj Quantum Information}\ }\textbf {\bibinfo {volume} {6}},\ \bibinfo {pages} {12} (\bibinfo {year} {2020})}\BibitemShut {NoStop}%
\bibitem [{\citenamefont {Bose}\ \emph {et~al.}(2025)\citenamefont {Bose}, \citenamefont {Fuentes}, \citenamefont {Geraci}, \citenamefont {Khan}, \citenamefont {Qvarfort}, \citenamefont {Rademacher}, \citenamefont {Rashid}, \citenamefont {Toro\ifmmode~\check{s}\else \v{s}\fi{}}, \citenamefont {Ulbricht},\ and\ \citenamefont {Wanjura}}]{Bose_2025}%
  \BibitemOpen
  \bibfield  {author} {\bibinfo {author} {\bibfnamefont {S.}~\bibnamefont {Bose}}, \bibinfo {author} {\bibfnamefont {I.}~\bibnamefont {Fuentes}}, \bibinfo {author} {\bibfnamefont {A.~A.}\ \bibnamefont {Geraci}}, \bibinfo {author} {\bibfnamefont {S.~M.}\ \bibnamefont {Khan}}, \bibinfo {author} {\bibfnamefont {S.}~\bibnamefont {Qvarfort}}, \bibinfo {author} {\bibfnamefont {M.}~\bibnamefont {Rademacher}}, \bibinfo {author} {\bibfnamefont {M.}~\bibnamefont {Rashid}}, \bibinfo {author} {\bibfnamefont {M.}~\bibnamefont {Toro\ifmmode~\check{s}\else \v{s}\fi{}}}, \bibinfo {author} {\bibfnamefont {H.}~\bibnamefont {Ulbricht}},\ and\ \bibinfo {author} {\bibfnamefont {C.~C.}\ \bibnamefont {Wanjura}},\ }\bibfield  {title} {\bibinfo {title} {Massive quantum systems as interfaces of quantum mechanics and gravity},\ }\href {https://doi.org/10.1103/RevModPhys.97.015003} {\bibfield  {journal} {\bibinfo  {journal} {Rev. Mod. Phys.}\ }\textbf {\bibinfo {volume} {97}},\ \bibinfo {pages} {015003} (\bibinfo {year}
  {2025})}\BibitemShut {NoStop}%
\bibitem [{\citenamefont {Hansen}\ \emph {et~al.}(2025)\citenamefont {Hansen}, \citenamefont {Minniberger}, \citenamefont {Dominik}, \citenamefont {Asenbaum}, \citenamefont {Higgins}, \citenamefont {Povey}, \citenamefont {Schmidt}, \citenamefont {Hofer}, \citenamefont {Claessen}, \citenamefont {Aspelmeyer},\ and\ \citenamefont {Trupke}}]{data_hansen_2025}%
  \BibitemOpen
  \bibfield  {author} {\bibinfo {author} {\bibfnamefont {J.}~\bibnamefont {Hansen}}, \bibinfo {author} {\bibfnamefont {S.}~\bibnamefont {Minniberger}}, \bibinfo {author} {\bibfnamefont {I.}~\bibnamefont {Dominik}}, \bibinfo {author} {\bibfnamefont {P.}~\bibnamefont {Asenbaum}}, \bibinfo {author} {\bibfnamefont {G.}~\bibnamefont {Higgins}}, \bibinfo {author} {\bibfnamefont {R.~G.}\ \bibnamefont {Povey}}, \bibinfo {author} {\bibfnamefont {P.}~\bibnamefont {Schmidt}}, \bibinfo {author} {\bibfnamefont {J.}~\bibnamefont {Hofer}}, \bibinfo {author} {\bibfnamefont {R.}~\bibnamefont {Claessen}}, \bibinfo {author} {\bibfnamefont {M.}~\bibnamefont {Aspelmeyer}},\ and\ \bibinfo {author} {\bibfnamefont {M.}~\bibnamefont {Trupke}},\ }\bibfield  {title} {\bibinfo {title} {Data used in the article "optical interferometric readout of a magnetically levitated superconducting microsphere"},\ }\href {https://doi.org/10.5281/zenodo.16779206} {10.5281/zenodo.16779206} (\bibinfo {year} {2025})\BibitemShut {NoStop}%
\bibitem [{\citenamefont {Cheng}\ \emph {et~al.}(2022)\citenamefont {Cheng}, \citenamefont {Liang}, \citenamefont {Kawamura}, \citenamefont {Zhou}, \citenamefont {Asamura}, \citenamefont {Uratani}, \citenamefont {Tiwari}, \citenamefont {Graham}, \citenamefont {Ohno}, \citenamefont {Nagai}, \citenamefont {Feng}, \citenamefont {Shigekawa},\ and\ \citenamefont {Cahill}}]{Cheng_2022}%
  \BibitemOpen
  \bibfield  {author} {\bibinfo {author} {\bibfnamefont {Z.}~\bibnamefont {Cheng}}, \bibinfo {author} {\bibfnamefont {J.}~\bibnamefont {Liang}}, \bibinfo {author} {\bibfnamefont {K.}~\bibnamefont {Kawamura}}, \bibinfo {author} {\bibfnamefont {H.}~\bibnamefont {Zhou}}, \bibinfo {author} {\bibfnamefont {H.}~\bibnamefont {Asamura}}, \bibinfo {author} {\bibfnamefont {H.}~\bibnamefont {Uratani}}, \bibinfo {author} {\bibfnamefont {J.}~\bibnamefont {Tiwari}}, \bibinfo {author} {\bibfnamefont {S.}~\bibnamefont {Graham}}, \bibinfo {author} {\bibfnamefont {Y.}~\bibnamefont {Ohno}}, \bibinfo {author} {\bibfnamefont {Y.}~\bibnamefont {Nagai}}, \bibinfo {author} {\bibfnamefont {T.}~\bibnamefont {Feng}}, \bibinfo {author} {\bibfnamefont {N.}~\bibnamefont {Shigekawa}},\ and\ \bibinfo {author} {\bibfnamefont {D.}~\bibnamefont {Cahill}},\ }\bibfield  {title} {\bibinfo {title} {High thermal conductivity in wafer-scale cubic silicon carbide crystals},\ }\href {https://doi.org/10.1038/s41467-022-34943-w} {\bibfield  {journal}
  {\bibinfo  {journal} {Nature Communications}\ }\textbf {\bibinfo {volume} {13}} (\bibinfo {year} {2022})}\BibitemShut {NoStop}%
\bibitem [{\citenamefont {Hofer}\ \emph {et~al.}(2023{\natexlab{b}})\citenamefont {Hofer}, \citenamefont {Gross}, \citenamefont {Higgins}, \citenamefont {Huebl}, \citenamefont {Kieler}, \citenamefont {Kleiner}, \citenamefont {Koelle}, \citenamefont {Schmidt}, \citenamefont {Slater}, \citenamefont {Trupke}, \citenamefont {Uhl}, \citenamefont {Weimann}, \citenamefont {Wieczorek},\ and\ \citenamefont {Aspelmeyer}}]{Hofer_2023}%
  \BibitemOpen
  \bibfield  {author} {\bibinfo {author} {\bibfnamefont {J.}~\bibnamefont {Hofer}}, \bibinfo {author} {\bibfnamefont {R.}~\bibnamefont {Gross}}, \bibinfo {author} {\bibfnamefont {G.}~\bibnamefont {Higgins}}, \bibinfo {author} {\bibfnamefont {H.}~\bibnamefont {Huebl}}, \bibinfo {author} {\bibfnamefont {O.~F.}\ \bibnamefont {Kieler}}, \bibinfo {author} {\bibfnamefont {R.}~\bibnamefont {Kleiner}}, \bibinfo {author} {\bibfnamefont {D.}~\bibnamefont {Koelle}}, \bibinfo {author} {\bibfnamefont {P.}~\bibnamefont {Schmidt}}, \bibinfo {author} {\bibfnamefont {J.~A.}\ \bibnamefont {Slater}}, \bibinfo {author} {\bibfnamefont {M.}~\bibnamefont {Trupke}}, \bibinfo {author} {\bibfnamefont {K.}~\bibnamefont {Uhl}}, \bibinfo {author} {\bibfnamefont {T.}~\bibnamefont {Weimann}}, \bibinfo {author} {\bibfnamefont {W.}~\bibnamefont {Wieczorek}},\ and\ \bibinfo {author} {\bibfnamefont {M.}~\bibnamefont {Aspelmeyer}},\ }\bibfield  {title} {\bibinfo {title} {High-$q$ magnetic levitation and control of superconducting microspheres
  at millikelvin temperatures},\ }\href {https://doi.org/10.1103/PhysRevLett.131.043603} {\bibfield  {journal} {\bibinfo  {journal} {Phys. Rev. Lett.}\ }\textbf {\bibinfo {volume} {131}},\ \bibinfo {pages} {043603} (\bibinfo {year} {2023}{\natexlab{b}})}\BibitemShut {NoStop}%
\bibitem [{\citenamefont {Neuhaus}\ and\ \citenamefont {et~al.}(2024)}]{PyRPL}%
  \BibitemOpen
  \bibfield  {author} {\bibinfo {author} {\bibfnamefont {L.}~\bibnamefont {Neuhaus}}\ and\ \bibinfo {author} {\bibnamefont {et~al.}},\ }\href@noop {} {\bibinfo {title} {Pyrpl}} (\bibinfo {year} {2024}),\ \bibinfo {note} {version 0.9.4.0 Available at: \url{https://pyrpl.readthedocs.io/en/latest/#}}\BibitemShut {NoStop}%
\bibitem [{\citenamefont {Polyanskiy}(2024)}]{Polyanskiy_2024}%
  \BibitemOpen
  \bibfield  {author} {\bibinfo {author} {\bibfnamefont {M.~N.}\ \bibnamefont {Polyanskiy}},\ }\bibfield  {title} {\bibinfo {title} {Refractiveindex.info database of optical constants},\ }\href {https://doi.org/10.1038/s41597-023-02898-2} {\bibfield  {journal} {\bibinfo  {journal} {Scientific Data}\ }\textbf {\bibinfo {volume} {11}},\ \bibinfo {pages} {94} (\bibinfo {year} {2024})}\BibitemShut {NoStop}%
\bibitem [{\citenamefont {Horowitz}\ \emph {et~al.}(1952)\citenamefont {Horowitz}, \citenamefont {Silvidi}, \citenamefont {Malaker},\ and\ \citenamefont {Daunt}}]{Horowitz_1952}%
  \BibitemOpen
  \bibfield  {author} {\bibinfo {author} {\bibfnamefont {M.}~\bibnamefont {Horowitz}}, \bibinfo {author} {\bibfnamefont {A.~A.}\ \bibnamefont {Silvidi}}, \bibinfo {author} {\bibfnamefont {S.~F.}\ \bibnamefont {Malaker}},\ and\ \bibinfo {author} {\bibfnamefont {J.~G.}\ \bibnamefont {Daunt}},\ }\bibfield  {title} {\bibinfo {title} {The specific heat of lead in the temperature range 1\ifmmode^\circ\else\textdegree\fi{}k to 75\ifmmode^\circ\else\textdegree\fi{}k},\ }\href {https://doi.org/10.1103/PhysRev.88.1182} {\bibfield  {journal} {\bibinfo  {journal} {Phys. Rev.}\ }\textbf {\bibinfo {volume} {88}},\ \bibinfo {pages} {1182} (\bibinfo {year} {1952})}\BibitemShut {NoStop}%
\bibitem [{\citenamefont {Chanin}\ and\ \citenamefont {Torre}(1972)}]{Chanin_1972}%
  \BibitemOpen
  \bibfield  {author} {\bibinfo {author} {\bibfnamefont {G.}~\bibnamefont {Chanin}}\ and\ \bibinfo {author} {\bibfnamefont {J.~P.}\ \bibnamefont {Torre}},\ }\bibfield  {title} {\bibinfo {title} {Critical-field curve of superconducting lead},\ }\href {https://doi.org/10.1103/PhysRevB.5.4357} {\bibfield  {journal} {\bibinfo  {journal} {Physical Review B}\ }\textbf {\bibinfo {volume} {5}},\ \bibinfo {pages} {4357} (\bibinfo {year} {1972})}\BibitemShut {NoStop}%
\bibitem [{\citenamefont {Tinkham}(2015)}]{Tinkham}%
  \BibitemOpen
  \bibfield  {author} {\bibinfo {author} {\bibfnamefont {M.}~\bibnamefont {Tinkham}},\ }\href@noop {} {\emph {\bibinfo {title} {Introduction to superconductivity}}},\ \bibinfo {edition} {2nd}\ ed.,\ Dover books on physics\ (\bibinfo  {publisher} {Dover Publ.},\ \bibinfo {address} {Mineola, NY},\ \bibinfo {year} {2015})\BibitemShut {NoStop}%
\bibitem [{\citenamefont {Ginsberg}(1962)}]{Ginsberg_1962}%
  \BibitemOpen
  \bibfield  {author} {\bibinfo {author} {\bibfnamefont {D.~M.}\ \bibnamefont {Ginsberg}},\ }\bibfield  {title} {\bibinfo {title} {Upper limit for quasi-particle recombination time in a superconductor},\ }\href {https://doi.org/10.1103/PhysRevLett.8.204} {\bibfield  {journal} {\bibinfo  {journal} {Physical Review Letters}\ }\textbf {\bibinfo {volume} {8}},\ \bibinfo {pages} {204} (\bibinfo {year} {1962})}\BibitemShut {NoStop}%
\bibitem [{\citenamefont {Kaplan}\ \emph {et~al.}(1976)\citenamefont {Kaplan}, \citenamefont {Chi}, \citenamefont {Langenberg}, \citenamefont {Chang}, \citenamefont {Jafarey},\ and\ \citenamefont {Scalapino}}]{Kaplan_1976}%
  \BibitemOpen
  \bibfield  {author} {\bibinfo {author} {\bibfnamefont {S.~B.}\ \bibnamefont {Kaplan}}, \bibinfo {author} {\bibfnamefont {C.~C.}\ \bibnamefont {Chi}}, \bibinfo {author} {\bibfnamefont {D.~N.}\ \bibnamefont {Langenberg}}, \bibinfo {author} {\bibfnamefont {J.~J.}\ \bibnamefont {Chang}}, \bibinfo {author} {\bibfnamefont {S.}~\bibnamefont {Jafarey}},\ and\ \bibinfo {author} {\bibfnamefont {D.~J.}\ \bibnamefont {Scalapino}},\ }\bibfield  {title} {\bibinfo {title} {Quasiparticle and phonon lifetimes in superconductors},\ }\href {https://doi.org/10.1103/PhysRevB.14.4854} {\bibfield  {journal} {\bibinfo  {journal} {Physical Review B}\ }\textbf {\bibinfo {volume} {14}},\ \bibinfo {pages} {4854} (\bibinfo {year} {1976})}\BibitemShut {NoStop}%
\bibitem [{\citenamefont {Hebestreit}\ \emph {et~al.}(2018)\citenamefont {Hebestreit}, \citenamefont {Frimmer}, \citenamefont {Reimann}, \citenamefont {Dellago}, \citenamefont {Ricci},\ and\ \citenamefont {Novotny}}]{Hebestreit_2018}%
  \BibitemOpen
  \bibfield  {author} {\bibinfo {author} {\bibfnamefont {E.}~\bibnamefont {Hebestreit}}, \bibinfo {author} {\bibfnamefont {M.}~\bibnamefont {Frimmer}}, \bibinfo {author} {\bibfnamefont {R.}~\bibnamefont {Reimann}}, \bibinfo {author} {\bibfnamefont {C.}~\bibnamefont {Dellago}}, \bibinfo {author} {\bibfnamefont {F.}~\bibnamefont {Ricci}},\ and\ \bibinfo {author} {\bibfnamefont {L.}~\bibnamefont {Novotny}},\ }\bibfield  {title} {\bibinfo {title} {Calibration and energy measurement of optically levitated nanoparticle sensors},\ }\href {https://doi.org/10.1063/1.5017119} {\bibfield  {journal} {\bibinfo  {journal} {Review of Scientific Instruments}\ }\textbf {\bibinfo {volume} {89}},\ \bibinfo {pages} {033111} (\bibinfo {year} {2018})}\BibitemShut {NoStop}%
\bibitem [{\citenamefont {Kubo}(1966)}]{Kubo_1966}%
  \BibitemOpen
  \bibfield  {author} {\bibinfo {author} {\bibfnamefont {R.}~\bibnamefont {Kubo}},\ }\bibfield  {title} {\bibinfo {title} {The fluctuation-dissipation theorem},\ }\href {https://doi.org/10.1088/0034-4885/29/1/306} {\bibfield  {journal} {\bibinfo  {journal} {Reports on Progress in Physics}\ }\textbf {\bibinfo {volume} {29}},\ \bibinfo {pages} {255} (\bibinfo {year} {1966})}\BibitemShut {NoStop}%
\bibitem [{\citenamefont {Povey}(2023)}]{Povey_Thesis}%
  \BibitemOpen
  \bibfield  {author} {\bibinfo {author} {\bibfnamefont {R.~G.}\ \bibnamefont {Povey}},\ }\emph {\bibinfo {title} {Two-Dimensional Optomechanical Resonators in Gallium Arsenide}},\ \href {https://doi.org/10.6082/uchicago.7543} {Ph.D. thesis},\ \bibinfo  {school} {The University of Chicago} (\bibinfo {year} {2023})\BibitemShut {NoStop}%
\end{thebibliography}%
\end{document}